\documentclass[trackchanges, twocolumn]{aastex7}

\usepackage{enumitem}
\usepackage{amsmath}
\usepackage{natbib}

\newcommand{\hh}{$\text{H}$}   
\newcommand{\he}{$\text{He}$}
\newcommand{\oo}{$\text{O}$}
\newcommand{\cc}{$\text{C}$}
\newcommand{\nn}{$\text{N}$}
\newcommand{\co}{$\text{CO}$}   

\newcommand{\heI}{$\text{He\,\textsc{i}}$}
\newcommand{\heII}{$\text{He\,\textsc{ii}}$}
\newcommand{\cII}{$\text{C\,\textsc{ii}}$}

\newcommand{\oI}{$\text{O\,\textsc{i}}$}
\newcommand{\ooI}{$[\text{O\,\textsc{i}}]$}
\newcommand{\oII}{$\text{O\,\textsc{ii}}$}
\newcommand{\ooII}{$[\text{O\,\textsc{ii}}]$}

\newcommand{\oIV}{$\text{O\,\textsc{iv}}$}
\newcommand{\oV}{$\text{O\,\textsc{v}}$}

\newcommand{\naI}{$\text{Na\,\textsc{i}}$}
\newcommand{\mgI}{$\text{Mg\,\textsc{i}}$}

\newcommand{\mgII}{$\text{Mg\,\textsc{ii}}$}
\newcommand{\siII}{$\text{Si\,\textsc{ii}}$}
\newcommand{\s}{$\text{S}$}
\newcommand{\caII}{$\text{Ca\,\textsc{ii}}$}

\newcommand{\kmsone}{$\text{km\ s}^{-1}$} 
\newcommand{\gcmthree}{$\text{g\ cm}^{-3}$} 
\newcommand{\KK}{$\text{K}\ $}    
\newcommand{\msun}{$\text{M}_{\odot}$}
\newcommand{\zsun}{$\text{Z}_{\odot}$}
\newcommand{\rot}{$\Omega / \Omega_{\text{crit}}$}
\begin{document}

\title{Rotational Dynamics in Pulsational Pair-Instability Supernovae: Implications for Mass-Loss and Transient Events}

\author[orcid=0009-0002-9359-1219,sname='Huynh']{Trang N. Huynh}
\affiliation{Department of Physics and Astronomy, Louisiana State University, Baton Rouge, LA, USA}
\email[show]{thuyn27@lsu.edu}  

\author[orcid=0000-0002-8179-1654,gname=Emmanouil, sname='Chatzopoulos']{Emmanouil Chatzopoulos} 
\affiliation{Department of Physics and Astronomy, Louisiana State University, Baton Rouge, LA, USA}
\affiliation{Institute of Astrophysics, Foundation for Research and Technology-Hellas (FORTH), Heraklion, 70013, Greece}
\email{chatzopoulos@lsu.edu}

\author[orcid=0009-0006-6787-461X]{Nageeb Zaman}
\affiliation{Department of Physics and Astronomy, Louisiana State University, Baton Rouge, LA, USA}
\email{mzama14@lsu.edu}

\begin{abstract}

    Pulsational pair-instability supernovae (PPISNe) are transient events occurring in progenitor stars with helium cores of  $\sim$32-65 \msun, where rapid electron-positron pair production induces pressure loss, collapse, and pulsations driving episodic mass loss. The number, strength, and duration of these pulses can lead to shell collisions that produce shock-powered transients, potentially explaining some of the most luminous events, such as superluminous supernovae, and other rare transients. Rapid progenitor rotation lowers the PPISN mass threshold and influences the dynamics, energetics, and chemical composition of PPISN-driven pulses. In this study, we computed 1D evolutionary models of massive, rotating PPISN progenitor stars with ZAMS masses of 85–140 \msun\ and metallicities of \zsun\ and 0.1 \zsun. Our analysis reveals strong correlations between PPISN ejected mass and total energy as well as between ejected mass and peak ejected shell velocity. Additionally, moderate correlations indicate that higher initial PPISN progenitor mass leads to greater mass ejection and energy release, while negative correlations show that rapid rotation appears to reduce mass ejection and kinetic energy of the shells. Subsequent pulses lead to hydrogen-poor, carbon- and oxygen-enriched ejected shells, indicating the effect of rotationally-induced chemical mixing PPISN-driven episodic mass-loss with implications for their transients. We model the light curve and synthetic spectra that arise from the collision of two \hh-poor shells for one of our models using the radiation transport code \texttt{SuperLite}. We find that shock-heated \hh-poor PPISN shell collisions from rapidly rotating progenitors can lead to moderately-luminous \hh-poor transients that share some similarities with observed SLSN-I events.

\end{abstract}

\section{INTRODUCTION} \label{sec:intro}

Massive stars have diverse fates. The interplay between their mass, metallicity, and rotation rate determines their long-term evolution, mass-loss, and terminal supernova (SN) explosion. Non-rotating stars within ZAMS masses of $130 - 260$ \msun\ experience violent dynamical instabilities. These instabilities arise during their evolution to the oxygen-burning phase, where the cores of these stars reach high central temperatures ($\sim1.5-2.2 \times 10^9$\KK) at relatively normal densities ($20,000-500,000$ \gcmthree), conditions that favor the rapid production of electron-positron pairs (\citealt{1964ApJS....9..201F}; \citealt{1967PhRvL..18..379B}; \citealt{1968Ap&SS...2...96F}; \citealt{2017ApJ...836..244W}). The contraction of the stellar core arises from the softening of the equation of state (EOS) combined with the slow-burning process of oxygen (\oo). This raises the central density and temperature to levels sufficient for the nuclear burning of carbon (\cc) and \oo, potentially producing large amounts of radioactive nickel-56 ($^{56}\text{Ni}$). This process can trigger an explosion that may result in a complete disruption of the star, known as pair-instability supernovae (PISNe), which have been extensively investigated (\citealt{1983A&A...119...61O}; \citealt{1984ApJ...280..825B}; \citealt{1985A&A...149..413G}; \citealt{2002ApJ...567..532H}; \citealt{2002ApJ...565..385U}; \citealt{2005ApJ...633.1031S}; \citealt{2011ApJ...734..102K}).

In contrast, stars with ZAMS masses between $95 - 130$ \msun\ in the non-rotating regime lack adequate energy to destroy the star (\citealt{2017ApJ...836..244W}). Instead, a series of pulses from the core violently eject solar masses of materials into the surrounding medium. Each pulse can eject material of different properties into the circumstellar medium in terms of ejected mass, kinetic energy, velocity, and composition, enriching the environment around the progenitor star. After the initial shell is expelled, the core shrinks until it achieves hydrostatic equilibrium, eventually reaching a central temperature of $10^{9}$ \KK, leading to another ejection. These cycles, which can span from a few days to millennia (\citealt{2007Natur.450..390W}), may also lead to collisions between ejected shells, depending on the initial conditions of the progenitor star. If the latter pulse travels faster, it catches up with the earlier one, resulting in shell-to-shell collision. Such interactions have the capacity to produce luminous transients, including superluminous supernovae (SLSNe) if the right conditions are met. This phenomenon is known as pulsational pair-instability supernovae or PPISNe.
 
Although not all PPISN light curves are luminous, their models have been invoked to match the observed light curves of SLSNe such as SN 2006gy (\citealt{2007Natur.450..390W}; \citealt{2007ApJ...666.1116S}), SN 2005ap (\citealt{2007ApJ...668L..99Q}), PTF15dam (\citealt{2017ApJ...835..266T}), and iPTF14hls (\citealt{2018ApJ...863..105W}). PPISN model based on the CSM interaction may also help interpret the light curves of Fast Blue Optical Transients (FBOTs) such as AT2018cow (\citealt{2020ApJ...903...66L}) and Fast Evolving Luminous Transients (FELTs) like KSN 2015K (\citealt{2018NatAs...2..307R}; \citealt{2019ApJ...881...35T}), whose luminosities are as bright as SLSNe but with a rapid rise and fall after peak luminosity. In addition, PPISN models have been involved to explain ultra-long Gamma-Ray Bursts (GRBs) in hydrogen-free (\hh-free) rotating progenitors, such as SN 2011kl (\citealt{2020A&A...641L..10M}).
 
Despite significant progress in the past two decades, a thorough analysis of the rotational dynamics of PPISNe emanating from fast-rotating progenitors is still missing. By examining the minimum mass limit to encounter pair-instability (PI) for zero-metallicity rotators, \citet{2012ApJ...748...42C} concluded that a lower Main-Sequence (MS) mass range with sufficiently strong rotationally induced mixing can encounter pulsational pair-instability (PPI) as low as 45 \msun. Higher MS masses lead to more massive helium (\he) cores, dominated by \cc, nitrogen (\nn), and \oo\ in the radiative envelope (\citealt{1985ApJ...292..506S}, \citealt{2000ApJ...528..368H}; \citealt{2000ARA&A..38..143M}). The ejected material from rapidly rotating PPISN progenitors can have different properties than in the case of no rotation. Namely, it can be \hh-poor and rich in \he, \cc, and \oo\ (\citealt{2012ApJ...760..154C}). An investigation of metallicity dependence on PPI events by \citet{2019ApJ...887...72L} shows that metallicity of at most 0.5 \zsun is required for stars to encounter PPI with \he-core masses greater than 40 \msun, as intense mass loss via stellar winds limits the formation of massive \hh\ cores.

This study presents a grid of models for 10\% \zsun  progenitors ranging from $85 - 140$ \msun\ that encounter PPI with various rotational rates. Some solar metallicity PPISN models are also explored, for comparison and to better understand the effects of metallicity in rotating PPISN dynamics. Section \ref{sec:mod} outlines the model setup and the physics of the simulations. Section \ref{sec:analysis} examines the characteristics of individual pulses and mass ejections in different rotational regimes and compares the results with those of previous studies. Sections \ref{sec:lightcurve} and \ref{sec:spec} present the light curve and spectra of two \hh-free shell collisions. Finally, Section \ref{sec:dis} summarizes our findings and discusses future work.

\section{METHODS} \label{sec:mod}  

In this section, we explore the rotational dynamics of the PPISN progenitors and implications for transient events through the following numerical setup. First, we simulate the evolution of PPISN progenitors using a one-dimensional stellar evolution code, Modules for Experiments in Stellar Astrophysics (\texttt{MESA}, \citealt{2011ApJS..192....3P, 2013ApJS..208....4P, 2015ApJS..220...15P, 2018ApJS..234...34P, 2019ApJS..243...10P}) to investigate how rotation affects the PPISN progenitors\footnote{The inlist files are available on Zenodo under an open-source Creative Commons Attribution license: \dataset[doi:10.5281/zenodo.15620230]{https://doi.org/10.5281/zenodo.15620230}.}. Then, we select a PPISN model of the collision of two \hh-poor shells and produce the bolometric luminosity using a radiation hydrodynamics code, \texttt{STELLA}. Finally, we generate synthetic spectra that arise from the shell collision through the radiation transport code \texttt{SuperLite} (\citealt{2023ApJ...953..132W}, \citealt{gururaj_wagle_2023_8111119}). The details of each setup are shown in the following subsections.

\subsection{PPISNe progenitor evolution with \texttt{MESA}}

To study the characteristics of PPISN progenitor stars and their pulses, we use the one-dimensional (1D) stellar evolution code, \texttt{MESA} (version r24.03.1). The models evolve from the Zero-Age Main Sequence (ZAMS) until the abundance of the \he core drops below the threshold $10^{-3}$, corresponding to a nominal phase of the \co core before encountering the PI. We then post-process the resulting PPISN models into an existing \texttt{MESA} test-suite setup called \texttt{ppisn}, which has been developed to showcase the capacity of \texttt{MESA} to model PPISNe from nonrotating, \hh-rich progenitor stars. The evolution of the model is followed from the start of the \cc-burning phase and through the occurrences of all PPI events to the terminal core-collapse (CC). For stellar models experiencing multiple pulses, we retain all gravitationally unbound ejecta and evolve them through subsequent pulses. No artificial truncation is performed.

In \texttt{MESA}, we can set the initial progenitor mass, metallicity, and rotation rate in terms of \rot, where $\Omega$ is defined as the value of the surface angular rotational velocity, and $\Omega_{\text{crit}}$ is the critical "break-up" rotational velocity, defined as $\Omega_{\text{crit}}^2 = (1 - \Gamma) G M / R^3$ in which $\Gamma = L/L_{edd}$ is the dimensionless Eddington factor, $L$ and $L_{edd}$ are the total radiated luminosity and the classical Eddington limit ($L_{edd} = 4\pi c G M / \kappa$), respectively, $G$ the gravitational constant, $c$ the speed of light, $M$ the mass, $R$ the radius of the star, and $\kappa$ the mass absorption coefficient \citep{1997ASPC..120..381L, 2019A&A...625A..88G}.

For the treatment of convective boundaries, we adopt the Ledoux criterion with the mixing length parameter $\alpha_{MLT} = 2.0$. We apply the efficiency of semi-convective mixing $\alpha_{S} = 1.0$ and without thermohaline mixing. Convection is turned off for locations with $v/c_s > 10$, where $c_s$ is the sound speed. We also include an exponential scheme to treat convective overshooting ~\citep*{2000A&A...360..952H} with parameters $f= 0.01$ and $f_0 = 0.005$. For the treatment of mass loss by stellar winds, our implementation follows that of ~\citet{2011A&A...530A.115B}. If the surface temperature is below $T_{\text{eff}}^{\text{jump}}$, a parameter formulated by ~\cite{2001A&A...369..574V} to account for the bi-stability jumps, we employ the mass loss rate as the maximum between the ~\cite{2000A&A...362..295V, 2001A&A...369..574V} and ~\cite{1990A&A...231..134N} rates. Otherwise, the mass-loss rate is a blend of ~\cite{2000A&A...362..295V, 2001A&A...369..574V} and ~\cite{1995A&A...299..151H}, weighted by hydrogen mass fraction, where the \cite{1995A&A...299..151H} wind is reduced by a factor of 10 for wind clumping effects. We turn off the wind mass loss during dynamical phases of the evolution, which means PPI mass ejections are the only source of mass loss mechanism in these phases. 

The hydrodynamics capabilities in \texttt{MESA} enable the modeling of pulses from the PPISNe as well as CC ejecta expansion. \texttt{MESA} current version uses a 1D, finite-volume Riemann solver with a Godunov-type scheme and an approximate HLLC (Harten-Lax-van Leer-Contact) Riemann solver, ensuring accurate shock capture and energy conservation without artificial viscosity. For PPISN modeling, the hydro solver activates when the following instability conditions are met: $1)$ the integrated $\langle \Gamma_1 \rangle - 4/3 < 0.01$, and 2) maximum velocity exceeds 50 \kmsone, where $\langle \Gamma_1 \rangle$ is the global weighted value of the adiabatic index, defined as: 
\begin{equation}
\langle \Gamma_1 \rangle = \frac{\int_{0}^{\text{M}} \frac{\Gamma_1\ \text{P}}{\rho} \text{dm}}{\int_{0}^{\text{M}} \frac{\text{P}}{\rho} \text{dm}}, 
\end{equation} 
where $\Gamma_1  = \left(\partial \ \ln \text{P} / \partial \ \ln\rho \right)|_{\text{ad}}$ is the local adiabatic index, M the total mass of the star, P and $\rho$ the local pressure and density, respectively. 

The solver remains active for 50 dynamical timescales (t$_{\text{dyn}}$), where t$_{\text{dyn}} = 1/\sqrt{G \langle \rho \rangle}$ with $\langle \rho \rangle = M / (4 \pi R^3 / 3)$, up to a mass fraction $q=0.9$. Before turning off the hydro solver, two more conditions must be met in order to allow more time for the additional shock to develop in the outer layers: the absolute velocities must be below 50 \kmsone, and the ratio of absolute velocity and sound speed must be below 0.5. Additionally, to ensure that the star is stabilizing before the hydro solver is off, neutrino and nuclear luminosities must be below $\log(L_{\text{neu}}/L_\odot) = 11$ and $\log(L_{\text{neu}}/L_\odot) = 10$. Further details on \texttt{MESA}'s hydrodynamics can be found in Section 4 of \citet{2018ApJS..234...34P}. 

By evolving a grid of PPISN progenitors in \texttt{MESA}, we obtain hydrodynamic profiles of the unbound and ejected materials after each pulse. The profiles provide insights into the rotational dynamics of the progenitors during PPI events. The two-shell collision of one model will be selected to produce the time evolution of the bolometric luminosity using \texttt{STELLA}.

\subsection{Radiation hydrodynamics with \texttt{STELLA}}

We model the light curve of the two-shell collision for one of our PPISN models using \texttt{STELLA}, a one-dimensional multigroup radiation hydrodynamic code embedded within \texttt{MESA} (\citealt{1998ApJ...496..454B}, \citealt{2000ApJ...532.1132B}, \citealt{2006A&A...453..229B}). \texttt{STELLA} implicitly solves time-dependent transfer equations for the angular moments of intensity in fixed frequency bins, coupled with the hydro equations in Lagrangian comoving grids. This allows to produce bolometric luminosities and broadband UBVRI colors. The combination of \texttt{MESA} and \texttt{STELLA} uses a 1D approximation on the evolution before and after the breakout that allows for less computational cost and thus produces useful results in a few hours \citep{2018ApJS..234...34P}. \texttt{STELLA} also provides zone-by-zone radiation hydrodynamics profiles in days post-breakout that give the input to \texttt{SuperLite} to generate the temporal evolution of the spectra. 

\texttt{STELLA} takes two input files of the hydrodynamic structures and composition from \texttt{MESA} and outputs the bolometric luminosities and velocities with respect to time. When transitioning from \texttt{MESA} to \texttt{STELLA}, we use the model profile when the materials are gravitationally unbound at the end of the first and second pulses. Under our assumption of homologous expansion, two pulses will then be run independently in \texttt{STELLA} so that the first pulse can propagate through the medium and eventually catch up with pulse two after a time $\Delta\text{t}_{\text{pulse}}$. We use the combination of the profile of pulse one at the time of pulse two ejections and the hydrodynamic profile of pulse two as \texttt{STELLA} inputs to produce the time evolution of the total bolometric light curve of the collision. The result will be detailed in Section \ref{sec:lightcurve}.

\subsection{Radiation transport with SuperLite}

\texttt{SuperLite} is a 1D, multi-group, Monte Carlo radiative transfer code that post-processes a single snapshot profile obtained from any hydrodynamic evolution code and produces the temporal evolution of the spectra. The code applies Implicit Monte Carlo and Discrete Diffusion Monte Carlo (IMC--DDMC) methods to model the radiation transport process in explosive and interacting outflows. \texttt{SuperLite} is an extension of the 123D homologous IMC--DDMC radiation transport code \texttt{SuperNu} \citep{2021ascl.soft03019W,2014ApJS..214...28W}, which has been used extensively to model spectra of regular SNe Type Ia and core--collapse SNe as well as other transients, such as kilonovae \citep{2024ApJ...961....9F}. 

We use the hydrodynamics profiles produced from \texttt{STELLA} from 10 days before to 30 days after the peak luminosity and truncate the profile at the Rosseland mean optical depth of 100, which practically includes all of the unbound mass following each PPISN pulse. The input profiles contain the shell collision's stratification information, including the physical quantities such as cell positions and velocities in radial coordinates, mass, volume, average temperature, and abundances. For the \texttt{SuperLite} radiation transport simulations, we set the number of source particles to be about one million ($2^{20}$) for better statistics and MC noise reduction and 6000 wavelength groups with a wavelength range of 1-30000~$\text{\AA}$ for opacity and output spectrum. We calculate the spectra assuming non-local thermodynamic equilibrium (NLTE) treatment for the rate matrix calculation of the excited level populations of \hh\ (see \citealt{2023ApJ...953..132W}). The resulting synthetic spectra are discussed in Section \ref{sec:spec}.

\section{ANALYSIS} \label{sec:analysis}

In this section, we present the results of our simulation of the evolution of the PPISN progenitors. The initial ZAMS masses for Z = 10\% \zsun\ range from 85-160 \msun\ for different degrees of rotation (non-rotating and rotating cases for \rot $= 0.2,\ 0.5,$ and 0.7). For models with high angular velocities (\rot $= 0.7$), we note that no part of the star reaches $\Omega > \Omega_{\text{crit}}$ or surface velocities exceeding the sound speed. Surface rotation remains subsonic, and rotational stabilities are treated within the \texttt{MESA} framework. We have also run a few models at solar metallicity (\zsun $=0.012$) and compare the properties of the ejected materials as a function of initial metallicity. For the solar metallicity models, we selected models with M$_{\text{ZAMS}}$ = 120, 140, and 160 \msun. This is a higher PPISN progenitor mass range than the one we used in the lower metallicity models since the solar metallicity models will experience more wind-driven substantial mass loss and, therefore, require a higher initial mass to enter the PI regime. The selected ZAMS range for solar metallicity models also leads to final CO core masses consistent with those investigated by \citet{2017ApJ...836..244W}, allowing us to compare our respective findings. The following subsections explore the effects of rotation, initial mass, and metallicity on PPISNe, discuss the pulse properties across parameter space, and compare our findings with non-rotating PPISN models.

\centerwidetable
\begin{deluxetable*}{ccccccccccccc} 
\tablenum{1}
\label{tab:lowZ}
\tablecaption{Characteristics of the 10\% \zsun Models}
\tablewidth{\textwidth}  
\tablehead{
\colhead{M$_{\text{ZAMS}}$} & \colhead{\rot} & \colhead{M$_{\text{He, dep}}$} & \colhead{M$_{\text{CO}}$} & \colhead{M$_{\text{total,eject}}$} & \colhead{M$_{\text{final}}$} & \colhead{No. of Pulses} & \colhead{PPI Duration} & \colhead{$\Delta \text{t}_{\text{final-CC}}$} & \colhead{E$_{\text{tot, max}}$} & \colhead{Fate\tablenotemark{a} } \\
\colhead{( \msun)} & \colhead{} & \colhead{( \msun)} & \colhead{( \msun)} & \colhead{( \msun)} & \colhead{( \msun)} & \colhead{} & \colhead{($10^{7}$ \text{s})} & \colhead{\((10^{7}\text{s})\)} & \colhead{\((10^{50} \text{erg})\)} & \colhead{}}
\startdata 
85 & 0.0 & 42.57 & 28.89 & $\_$ & $\_$ & $\_$ & $\_$ & $\_$ & $\_$ & CC \\ 
85 & 0.2 & 59.71 & 54.70 & 7.46 & 51.98 & 2 & 1864.48 & 0.34 & 1.41 & PPISN \\ 
85 & 0.5 & 53.27 & 50.58 & 3.85 & 48.96 & 3 & 0.71 & 0.49 & 0.92 & PPISN \\ 
85 & 0.7 & 52.50 & 49.68 & 2.95 & 48.81 & 3 & 0.25 & 0.08 & 0.68 & PPISN \\ 
90 & 0.0 & 45.72 & 31.30 & $\_$ & $\_$ & $\_$ & $\_$ & $\_$ & $\_$ & CC \\ 
90 & 0.2 & 61.01 & 55.96 & 9.78 & 50.95 & 2 & 6119.90 & 1.01 & 1.05 & PPISN \\ 
90 & 0.5 & 55.92 & 52.66 & 4.99 & 50.43 & 2 & 1.84 & 42.49 & 0.64 & PPISN \\ 
90 & 0.7 & 54.82 & 52.01 & 3.96 & 50.16 & 2 & 0.40 & 3.85 & 0.79 & PPISN \\ 
95 & 0.0 & 48.62 & 33.78 & $\_$ & $\_$ & $\_$ & $\_$ & $\_$ & $\_$ & CC \\ 
95 & 0.2 & 64.43 & 59.13 & 9.08 & 54.75 & 1 & 0.03 & 12404.58 & 3.85 & PPISN \\ 
95 & 0.5 & 58.48 & 54.59 & 4.55 & 53.49 & 2 & 0.24 & 355.39 & 0.78 & PPISN \\ 
95 & 0.7 & 57.02 & 54.45 & 5.27 & 51.06 & 2 & 1.98 & 363.03 & 0.68 & PPISN \\ 
110 & 0.0 & 57.13 & 40.47 & 6.24 & 46.12 & 1 & 0.07 & 12843.68 & 0.84 & PPISN \\ 
T110B & 0.0 & 49.50 & 44.67 & 39.40 & 44.70 & 2 & 9.50 & $\_$ & 7.40 & PPISN \\ 
110 & 0.2 & 74.18 & 68.43 & $\_$ & $\_$ & $\_$ & $\_$ & $\_$ & $\_$ & PISN \\ 
110 & 0.5 & 66.33 & 62.43 & 13.49 & 52.21 & 2 & 203.63 & 11294.31 & 3.77 & PPISN \\ 
110 & 0.7 & 64.66 & 61.35 & $\_$ & $\_$ & $\_$ & $\_$ & $\_$ & $\_$ & CC \\ 
115 & 0.0 & 60.44 & 42.77 & $\_$ & $\_$ & $\_$ & $\_$ & $\_$ & $\_$ & CC \\ 
115 & 0.2 & 78.20 & 72.03 & $\_$ & $\_$ & $\_$ & $\_$ & $\_$ & $\_$ & PISN \\ 
115 & 0.5 & 68.61 & 64.31 & 24.23 & 43.70 & 2 & 5063.13 & 14965.09 & 5.30 & PPISN \\ 
115 & 0.7 & 67.28 & 63.30 & 23.42 & 42.99 & 2 & 57.18 & 16590.74 & 2.70 & PPISN \\ 
120 & 0.0 & 62.97 & 45.18 & $\_$ & $\_$ & $\_$ & $\_$ & $\_$ & $\_$ & CC \\ 
120 & 0.2 & 80.52 & 74.35 & $\_$ & $\_$ & $\_$ & $\_$ & $\_$ & $\_$ & PISN \\ 
120 & 0.7 & 69.19 & 65.38 & $\_$ & $\_$ & $\_$ & $\_$ & $\_$ & $\_$ & PISN \\ 
T123A & 0.0 & 55.79 & 50.38 & 24.20 & 50.20 & 2 & 3900 & $\_$ & 17.00 & PPISN \\ 
130 & 0.0 & 69.00 & 50.27 & $\_$ & $\_$ & $\_$ & $\_$ & $\_$ & $\_$ & CC \\ 
T130D & 0.0 & 59.96 & 54.28 & 91.20 & 38.80 & 3 & 25000.00 & $\_$ & 41.00 & PPISN \\ 
130 & 0.2 & 84.09 & 77.93 & $\_$ & $\_$ & $\_$ & $\_$ & $\_$ & $\_$ & PISN \\ 
130 & 0.5 & 75.99 & 70.85 & $\_$ & $\_$ & $\_$ & $\_$ & $\_$ & $\_$ & PISN \\ 
130 & 0.7 & 74.04 & 69.45 & $\_$ & $\_$ & $\_$ & $\_$ & $\_$ & $\_$ & PISN \\ 
140 & 0.0 & 74.26 & 55.03 & $\_$ & $\_$ & $\_$ & $\_$ & $\_$ & $\_$ & PISN \\ 
140 & 0.2 & 89.48 & 82.85 & $\_$ & $\_$ & $\_$ & $\_$ & $\_$ & $\_$ & PISN \\ 
140 & 0.7 & 78.21 & 74.05 & $\_$ & $\_$ & $\_$ & $\_$ & $\_$ & $\_$ & PISN \\ 
160 & 0.0 & 85.89 & 64.39 & $\_$ & $\_$ & $\_$ & $\_$ & $\_$ & $\_$ & PISN \\ 
160 & 0.2 & 99.16 & 91.94 & $\_$ & $\_$ & $\_$ & $\_$ & $\_$ & $\_$ & PISN \\ 
160 & 0.7 & 87.12 & 82.02 & $\_$ & $\_$ & $\_$ & $\_$ & $\_$ & $\_$ & PISN \\ 
\enddata
\tablenotetext{a}{CC: core collapse; PPISN: pulsational pair-instability supernova; PISN: pair-instability supernova.}
\tablecomments{Fates are determined from full simulation outcomes, with marginal cases sensitive to input physics (see Section \ref{sec:analysis})}. Model T110B, T123A, and T130D are the models T110B, T123A, and T130D shown in Table 2 of \citet{2017ApJ...836..244W}, where A, B, D show the standard mass loss rate multiplied by a factor of $1/2$, $1/4$, and 0, respectively.
\end{deluxetable*}

Table \ref{tab:lowZ} summarizes the properties of the PPISN progenitor star models explored in this work. From left to right, the columns represent the following: M$_{\text{ZAMS}}$, the ZAMS mass of the models; \rot, the critical rotational ratio as defined in Section \ref{sec:mod}; M$_{\text{He, dep}}$ and \co, the \he- and \co-core masses at the point where the central \he\ mass fraction falls below $10^{-3}$; M$_{\text{total,eject}}$, {\it No. of Pulses}, and {\it PPI Duration} are the total mass ejected, number of pulses, and total duration from the onset of the first pulse to the end of the last pulse; $\Delta \text{t}_{\text{final-CC}}$ is the time between the final mass ejection and the onset of core collapse; E$_{\text{tot,max}}$ is the maximum total energy (potential, kinetic, and internal energy) corresponding to the strongest pulse; M$_{\text{final}}$, the final mass after pulsations; and the final fate of the star—classified as CC (core--collapse), PPISN, or full--fledged PISN (pair instability supernova). The fates are determined from the actual outcomes of the simulations, tracking the full evolution through all PPI phases of each model, rather than inferred solely from \co-core mass. In marginal cases, fates might be sensitive to input physics such as stellar evolution assumptions.

In comparison to the effect of rotation on pulse ejections, we select three non-rotating PPISNe models in \cite{2017ApJ...836..244W} at 10\% \zsun. They include Model T110B with $1/4$ standard mass loss rate, T123A with $1/2$ standard mass loss rate, and T130D with no mass loss. These models have \he-core masses similar to our rotating models: 85 \msun\ at \rot $= 0.7$, 90 \msun\ at \rot $= 0.5$ and $0.7$, and 90 \msun\ at \rot $= 0.2$, respectively.

\subsection{Rotational Effects on PPISNe with Fixed M\textsubscript{ZAMS}}

Comparing models with the same M$_{\text{ZAMS}}$ but different rotation rates, increased rotation generally leads to the following effects:
\begin{enumerate}[label=(\roman*),leftmargin=12pt,topsep=0pt,partopsep=0pt,parsep=1pt,itemsep=1pt]
    \item increased \he- and \co-core mass when the angular velocity is small ($\le 20\%\  \Omega_{\text{crit}}$) while reduced \co-core mass when the angular velocity is large ($20-70\%\ \Omega_{\text{crit}}$). The decrease in \co-core mass at high rotation is due to the combination effect of:
    \begin{enumerate}[label=(\alph*),leftmargin=8pt,topsep=0pt,partopsep=0pt,parsep=1pt,itemsep=1pt]
        \item enhanced rotational mixing, which dredges heavier elements (CNO) to the surface, increasing surface opacity and enabling stronger wind-driven mass loss, especially during the main sequence and core \he-burning phases;
        \item rotational boosting of mass loss directly through surface deformation and gravity darkening (e.g., \citealt{1997ASPC..120..381L})
    \end{enumerate}
    This leads to overall mass loss exceeding the rate at which rotational mixing builds up the core, resulting in a net decrease in final \co-core mass at higher \rot.
    \item decreased M$_{\text{eject}}$ due to higher rotation models exhibiting enhanced mass-loss rates from (a) higher surface metal abundance favoring stronger wind-driven mass loss and (b) mechanical effects of rotation boosting mass-loss rates, resulting in pulses from a smaller, more gravitationally bound star;
    \item an average of $\sim 2$ pulses across all cases, with the number of pulses fluctuating;
    \item decreased duration between pulses for lower masses ($85 - 90$ \msun) but consistent or slightly increased duration for higher masses, noting that high-mass PPISN models (123A and 130D) with suppressed mass-loss rates did not produce full-fledged PISN;
    \item models with \rot $> 0.5$ exhibiting less extended envelopes during the first pulse than those at $20\%$ critical rotation, as rapid rotation leads to a compact stellar structure, with centrifugal forces redistributing material, reducing gravitational binding, and enhanced rotational mixing supporting prolonged nuclear burning and core contraction, while gravity darkening and increased mass loss limit envelope expansion by reducing radiative flux and pressure support.
\end{enumerate}

\subsection{M$_{\text{ZAMS}}$ Effects on PPISNe at Fixed \texorpdfstring{\rot}{Omega / Omega crit}}

Comparing models with same rotation rate but different M$_{\text{ZAMS}}$:
\begin{enumerate}[label=(\roman*)]
    \item higher mass produces higher M$_{\text{CO}}$ (as expected);
    \item higher mass produces slightly less pulses because the earlier pulses are stronger and more energetic, thus removing more mass than models with lower M$_{\text{ZAMS}}$ (M$_{\text{total,eject}}$ is higher for higher M$_{\text{ZAMS}}$.)
\end{enumerate}

\subsection{Solar vs. 10\% Solar Metallicity Models}

In comparison between 10\% \zsun\ and \zsun\ progenitors, Table \ref{tab:solarZ} summarizes the solar metallicity models with the same properties as Table \ref{tab:lowZ}. Note that the fates of these models are also determined from the actual outcomes of the simulations, not solely inferred from \co-core mass. The 140 \msun\ and 160 \msun\ solar metallicity models with \rot $= 0.2$ produce PPISNe while the 10\% \zsun\ models of the same M$_{\text{ZAMS}}$ and rotation rate produce full-fledge PISNe. This difference arises from the effect of metallicity on wind-driven mass-loss during the evolution of progenitor stars: the 10\%~\zsun\ models suffer weaker mass-loss, retaining a higher \co-core mass. As a result, these progenitors undergo a stronger core implosion, leading to a terminal, full-fledge PISNe instead of pulsational behavior. The 120 \msun\ solar metallicity model, however, reaches a PPI region whose \co-core mass is too small to become a pulsational event, thus likely evolving to a core--collapse.

\begin{deluxetable*}{cccccccccccccc}
    \tablenum{2}
    \label{tab:solarZ}
    \tablecaption{Characteristics of the Z = \(Z_\odot\) Models}
    \tablewidth{20pt}
    \tablehead{
    \colhead{M$_{\text{ZAMS}}$} & \colhead{\rot} & \colhead{M$_{\text{He, dep}}$} & \colhead{M$_{\text{CO}}$} & \colhead{M$_{\text{total,eject}}$} & \colhead{M$_{\text{final}}$} & \colhead{No. of Pulses} & \colhead{PPI Duration} & \colhead{$\Delta \text{t}_{\text{final-CC}}$} & \colhead{E$_{\text{tot, max}}$} & \colhead{Fate\tablenotemark{a} } \\
    \colhead{( \msun)} & \colhead{} & \colhead{( \msun)} & \colhead{( \msun)} & \colhead{( \msun)} & \colhead{( \msun)} & \colhead{} & \colhead{($10^{7}$ \text{s})} & \colhead{\((10^{7}\text{s})\)} & \colhead{\((10^{50} \text{erg})\)} & \colhead{}}
    \startdata
    120 & 0.0 & 61.22 & 44.11 & $\_$ & $\_$ & $\_$ & $\_$ & $\_$ & $\_$ & PISN \\ 
    120 & 0.2 & 38.39 & 33.76 & $\_$ & $\_$ & $\_$ & $\_$ & $\_$ & $\_$ & PPISN \\ 
    120 & 0.7 & 32.85 & 28.72 & $\_$ & $\_$ & $\_$ & $\_$ & $\_$ & $\_$ & CC \\ 
    140 & 0.0 & 71.4 & 52.96 & $\_$ & $\_$ & $\_$ & $\_$ & $\_$ & $\_$ & PISN \\ 
    140 & 0.2 & 41.53 & 36.65 & 1.58 & 39.11 & 2.0 & 0.12 & 0.10 & 0.56 & PPISN \\ 
    140 & 0.7 & 36.24 & 31.95 & $\_$ & $\_$ & $\_$ & $\_$ & $\_$ & $\_$ & CC \\ 
    160 & 0.0 & 81.93 & 61.86 & $\_$ & $\_$ & $\_$ & $\_$ & $\_$ & $\_$ & PISN \\ 
    160 & 0.2 & 44.99 & 39.83 & 1.02 & 43.02 & 1.0 & 0.03 & 0.52 & 0.29 & PPISN \\ 
    160 & 0.7 & 39.01 & 34.37 & $\_$ & $\_$ & $\_$ & $\_$ & $\_$ & $\_$ & CC \\ 
    \enddata
    \tablenotetext{a}{ CC: core collapse; PPISN: pulsational pair-instability supernova; PISN: pair-instability supernova.}
    \tablecomments{Fates are determined from full simulation outcomes, with marginal cases sensitive to input physics (see Section \ref{sec:analysis})}.
\end{deluxetable*}

\subsection{Pulse Properties Across Mass-Rotation Space}

\begin{figure*}[h!t]
    \centering 
    \fig{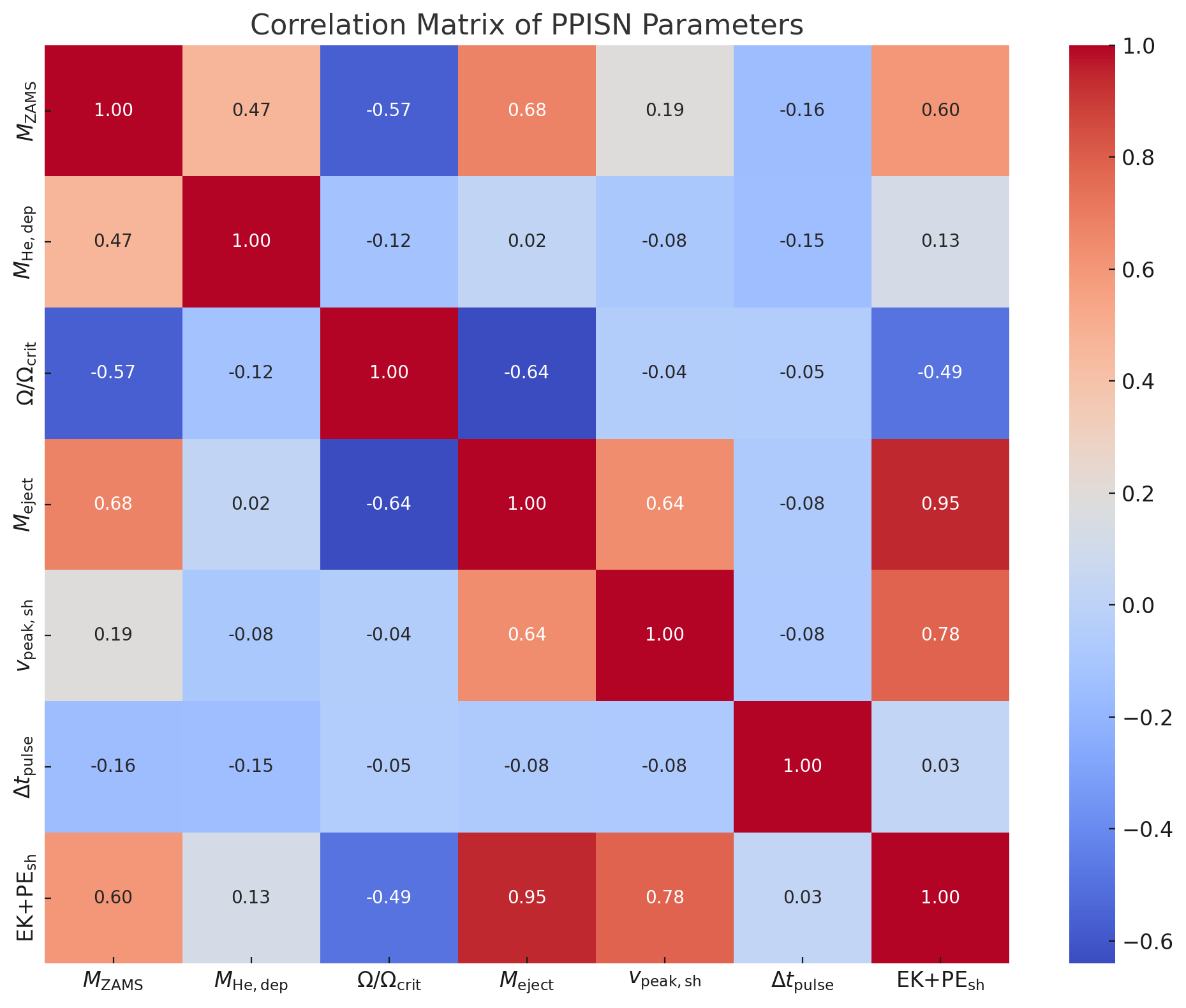}{0.7\textwidth}{} 
    \caption{Correlation matrix on the stellar parameters based on the 33 models presented in Table \ref{tab:lowZ}, showing the relationships between initial ZAMS mass (M$_{\text{ZAMS}}$), \he-core mass at depletion (M$_{\text{He,dep}}$), rotational rate (\rot), ejected mass (M$_{\text{eject}}$), peak shell velocity (v$_{\text{peak, sh}}$), the pulse duration ($\Delta \text{t}_{\text{pulse}}$), and the combined energy released ($\text{EK+PE}_{\text{sh}}$). Red indicates positive correlations while blue indicates negative correlation. \label{fig:corr}}
\end{figure*}

Table \ref{tab:shells} gives details of the ejected shells during individual pulses for all PPISN models at Z = 0.1 \zsun. The last two models, 140 and 160 \msun\ marked with an asterisk ($^*$), belong to the Z = \zsun\ PPISN progenitors. The first three columns show M$_{\text{ZAMS}}$, M$_{\text{He,dep}}$, \rot, as explained in the previous tables. The subsequent columns describe the characteristics of each pulse: M$_{\text{eject}}$, the mass ejected during that pulse; X$_{\text{He,sh}}$, X$_{\text{C,sh}}$ , and X$_{\text{O,sh}}$, the mass fractions of \he, \cc, and \oo\ in the shell; v$_{\text{peak,sh}}$ is the peak velocity of the ejected material. Also given is {\it Pulse Duration}, measured from the beginning to the end of the pulse, and $\Delta \text{t}_{\text{pulse}}$ represents the time between the end of one pulse to the start of the next. Note that in Table \ref{tab:shells},  $\Delta \text{t}_{\text{pulse}}$ values start from pulse two because they measure the time between consecutive pulses. By definition, the $\Delta \text{t}_{\text{pulse}}$ value for pulse one is left blank because there is no preceding pulse to measure. For example, in Model 85 \msun\ at \rot $= 0.5$ (three pulses), the $\Delta \text{t}_{\text{pulse}}$ value for pulse 2 is 7.19 days which is the time elapsed from the end of pulse 1 to the start of pulse 2. Similarly, $\Delta \text{t}_{\text{pulse}}$ value for pulse 3 is 19.66 days, the time between the end of pulse two and the start of pulse 3. Lastly, E$_{\text{tot,sh}}$ is the total energy (potential, kinetic, and internal energies) of the ejected material. 

\begin{deluxetable*}{cccccccccccc}
    \tablenum{3}
    \tablecaption{Physical Characteristics of the Ejected Shells by PPISN Models}
    \tablewidth{20pt}
    \label{tab:shells}
    \tablehead{
    \colhead{M$_{\text{ZAMS}}$} & \colhead{M$_{\text{He,dep}}$} & \colhead{\rot} 
    & \colhead{Pulse} & \colhead{M$_{\text{eject}}$} & \colhead{X$_{\text{He,sh}}$} & \colhead{X$_{\text{C,sh}}$} 
    & \colhead{X$_{\text{O,sh}}$} & \colhead{v$_{\text{peak,sh}}$} & \colhead{Pulse Duration} 
    & \colhead{$\Delta \text{t}_{\text{pulse}}$} & \colhead{E$_{\text{tot,sh}}$}\\ 
    \colhead{( \msun)} & \colhead{( \msun)} & \nocolhead{Name} & \nocolhead{Name} 
    & \colhead{( \msun)}  & \nocolhead{Name} & \nocolhead{Name} & \nocolhead{Name}
    & \colhead{\kmsone} & \colhead{(days)} & \colhead{(days)} & \colhead{($10^{50}$ erg)}}
    \startdata
    85 & 59.71 & 0.2 & 1 & 4.84 & 0.15 & 0.29 & 0.52 & 1413 & 2.75 & $\_$ & 0.58 \\ 
    $\_$ & $\_$ & $\_$ & 2 & 2.62 & 0.01 & 0.16 & 0.75 & 3350 & 0.11 & 215793.08 & 1.41 \\ 
    85 & 53.27 & 0.5 & 1 & 0.81 & 0.23 & 0.4 & 0.36 & 3412 & 1.35 & $\_$ & 0.39 \\ 
    $\_$ & $\_$ & $\_$ & 2 & 1.64 & 0.13 & 0.31 & 0.54 & 4086 & 0.26 & 13.78 & 0.92 \\ 
    $\_$ & $\_$ & $\_$ & 3 & 1.4 & 0.04 & 0.19 & 0.73 & 4017 & 0.27 & 66.76 & 0.85 \\ 
    85 & 52.5 & 0.7 & 1 & 0.71 & 0.17 & 0.39 & 0.43 & 3410 & 1.74 & $\_$ & 0.39 \\ 
    $\_$ & $\_$ & $\_$ & 2 & 1.24 & 0.13 & 0.34 & 0.53 & 3725 & 0.26 & 7.19 & 0.68 \\ 
    $\_$ & $\_$ & $\_$ & 3 & 1.0 & 0.08 & 0.27 & 0.63 & 3595 & 0.27 & 19.66 & 0.53 \\ 
    90 & 61.01 & 0.2 & 1 & 6.79 & 0.05 & 0.19 & 0.67 & 1260 & 4.56 & $\_$ & 0.82 \\ 
    $\_$ & $\_$ & $\_$ & 2 & 2.99 & 0.0 & 0.14 & 0.8 & 1980 & 0.35 & 708316.48 & 1.05 \\ 
    90 & 55.92 & 0.5 & 1 & 2.15 & 0.18 & 0.36 & 0.45 & 2498 & 5.65 & $\_$ & 0.6 \\ 
    $\_$ & $\_$ & $\_$ & 2 & 2.84 & 0.03 & 0.19 & 0.73 & 2655 & 1.37 & 205.67 & 0.64 \\ 
    90 & 54.82 & 0.7 & 1 & 1.58 & 0.16 & 0.36 & 0.47 & 3623 & 4.27 & $\_$ & 0.79 \\ 
    $\_$ & $\_$ & $\_$ & 2 & 2.38 & 0.04 & 0.21 & 0.71 & 2769 & 0.96 & 41.62 & 0.72 \\ 
    95 & 64.43 & 0.2 & 1 & 9.08 & 0.04 & 0.19 & 0.67 & 3171 & 2.95 & $\_$ & 3.85 \\ 
    95 & 58.48 & 0.5 & 1 & 0.78 & 0.24 & 0.41 & 0.34 & 2054 & 0.24 & $\_$ & 0.18 \\ 
    $\_$ & $\_$ & $\_$ & 2 & 3.77 & 0.1 & 0.26 & 0.62 & 2126 & 1.48 & 25.97 & 0.78 \\ 
    95 & 57.02 & 0.7 & 1 & 2.34 & 0.12 & 0.28 & 0.58 & 2676 & 7.99 & $\_$ & 0.68 \\ 
    $\_$ & $\_$ & $\_$ & 2 & 2.94 & 0.01 & 0.15 & 0.78 & 2432 & 1.93 & 218.85 & 0.62 \\ 
    110 & 57.12 & 0.0 & 1 & 6.24 & 0.01 & 0.05 & 0.75 & 1493 & 8.32 & $\_$ & 0.84 \\ 
    110B & 49.5 & 0.0 & 1 & 35.3 & $\_$ & $\_$ & $\_$ & 1300 & $\_$ & $\_$ & 5.1 \\ 
    $\_$ & $\_$ & $\_$ & 2 & 4.1 & $\_$ & $\_$ & $\_$ & 2000-3500 & $\_$ & 986.0 & 2.3 \\ 
    110 & 66.33 & 0.5 & 1 & 4.14 & 0.15 & 0.29 & 0.52 & 3364 & 0.51 & $\_$ & 1.57 \\ 
    $\_$ & $\_$ & $\_$ & 2 & 9.35 & 0.0 & 0.11 & 0.79 & 3300 & 2.59 & 64.56 & 3.77 \\ 
    115 & 68.61 & 0.5 & 1 & 7.17 & 0.02 & 0.15 & 0.75 & 1079 & 5.12 & $\_$ & 0.87 \\ 
    $\_$ & $\_$ & $\_$ & 2 & 17.06 & 0.0 & 0.08 & 0.8 & 2508 & 7.54 & 1605.47 & 5.3 \\ 
    115 & 67.28 & 0.7 & 1 & 3.6 & 0.15 & 0.32 & 0.51 & 3154 & 0.44 & $\_$ & 1.34 \\ 
    $\_$ & $\_$ & $\_$ & 2 & 19.82 & 0.0 & 0.09 & 0.79 & 1540 & 15.56 & 18.09 & 2.7 \\ 
    123A & 55.79 & 0.0 & 1 & 17.8 & $\_$ & $\_$ & $\_$ & 1000-3000 & 120 & $\_$ & 11.0 \\ 
    $\_$ & $\_$ & $\_$ & 2 & 6.4 & $\_$ & $\_$ & $\_$ & $\_$ & $\_$ & 368650 & 6.0 \\ 
    130D & 59.96 & 0.0 & 1 & 70.0 & $\_$ & $\_$ & $\_$ & $\_$ & $\_$ & $\_$ & 15.0 \\ 
    $\_$ & $\_$ & $\_$ & 2 & 7.7 & $\_$ & $\_$ & $\_$ & $\_$ & $\_$ & 1204500 & 11.0 \\ 
    $\_$ & $\_$ & $\_$ & 3 & 13.5 & $\_$ & $\_$ & $\_$ & $\_$ & $\_$ & 240 & 15.0 \\ 
    140* & 41.53 & 0.2 & 1 & 0.37 & 0.41 & 0.45 & 0.13 & 1655 & 0.18 & $\_$ & 0.06 \\ 
    $\_$ & $\_$ & $\_$ & 2 & 1.2 & 0.36 & 0.46 & 0.16 & 4012 & 0.67 & 13.13 & 0.56 \\ 
    160* & 44.99 & 0.2 & 1 & 1.02 & 0.37 & 0.46 & 0.15 & 2449 & 3.92 & $\_$ & 0.29 \\ 
    \enddata
\end{deluxetable*}

According to Table \ref{tab:shells}, at \rot $= 0.2$, PPISN models (85 \msun, 90 \msun, 95 \msun) typically produce two mass ejections. In Model 90 \msun, the first pulse ejects the highest mass, characterized by a peak velocity of $\sim$ 1300 \kmsone and occurs over a short timescale ($\sim$ 5 days). There is a quiescent period of 2000 years when the star is in hydrostatic equilibrium. A second eject of less than a solar mass happens, traveling at a peak velocity of $\sim$2000 \kmsone. The second pulse exhibits a slightly higher energy of 3.6 $\times 10^{49}$ erg and is predominantly \oo-rich, with moderate \cc\ and less than 15\% \he. These rotating models also lack \hh-envelope due to mass loss via stellar winds. 

At \rot $= 0.5$, mass ejections differs significantly. All models in this regime undergo two or three pulses, with the second pulse generally exhibiting the highest kinetic energy and peak velocity. Increased rotation fosters chemical homogeneity, as evident in the \he, \cc, and \oo\ mass fractions of the ejected shells. In the first pulse, hydrogen constitutes 2\%-24\% of the ejected mass, dropping below 10\% in subsequent pulses. Oxygen dominates the second ejection, comprising more than 60\% of the shell, consistent with \cite{2012ApJ...760..154C}. 


We conduct a correlation analysis on the stellar parameters shown in Figure \ref{fig:corr} to quantify the relationships among them, based on the full suite of 33 models described in Table \ref{tab:lowZ}. This sample size, while moderate, encompasses a diverse set of progenitor models spanning a range of initial masses (M$_{\text{ZAMS}}$), rotation rates (\rot), and pulse dynamics, allowing us to capture key trends in the parameter space. Here we show the following key trends from the matrix:

\begin{itemize}[leftmargin=0pt,labelwidth=0pt,align=left]
    \item \textbf{Strong positive correlations:} 
    \begin{itemize}[leftmargin=3pt,labelwidth=0pt,align=left]
        \renewcommand{\labelitemii}{\(\circ\)}
        \item M$_{\text{eject}}$ and E$_{\text{KE,sh}}$ (r$\approx$0.95): The amount of ejected mass is strongly correlated with the kinetic energy of the shell, suggesting that higher mass ejection contributes to significant energy release.
        \item M$_{\text{eject}}$ and v$_{\text{peak,sh}}$ (r$\approx$0.64): Larger ejected mass correlates with higher peak velocities of the ejected material.
    \end{itemize}
    \item \textbf{Moderate positive correlations:}
    \begin{itemize}[leftmargin=3pt,labelwidth=0pt,align=left]
        \renewcommand{\labelitemii}{\(\circ\)}
        \item M$_{\text{ZAMS}}$ and M$_{\text{eject}}$ (r$\approx$0.68): Higher initial stellar mass tends to produce greater mass ejection during the explosion.
        \item M$_{\text{ZAMS}}$ and E$_{\text{KE,sh}}$ (r$\approx$0.60): Higher initial stellar mass is associated with greater kinetic energy release.
    \end{itemize}
    \item \textbf{Negative correlations:}
    \begin{itemize}[leftmargin=3pt,labelwidth=0pt,align=left]
        \renewcommand{\labelitemii}{\(\circ\)}
        \item \rot\ and M$_{\text{eject}}$ (r$\approx$-0.64): Rapid rotation appears to reduce the amount of mass ejection, possibly due to angular momentum effects stabilizing the core. 
        \item \rot\ and E$_{\text{KE,sh}}$ (r$\approx$-0.49): Faster rotation is associated with lower energy release.
    \end{itemize}
    \item \textbf{Weak or no significant correlations:}
    \begin{itemize}[leftmargin=3pt,labelwidth=0pt,align=left]
        \renewcommand{\labelitemii}{\(\circ\)}
        \item $\text{Pulse\ Duration}$ shows weak or negligible correlations with most parameters, suggesting its variability may depend on factors not strongly represented in this dataset.
    \end{itemize}
\end{itemize}

The correlation matrix in Figure \ref{fig:corr} provides a quantitative overview of relationships between progenitor properties and explosion characteristics. Physically, the strong correlation between ejected mass and kinetic energy (r $\approx0.95$) reflects conservation of energy: more massive shells require more energy to be ejected at comparable velocities. The correlation between ejected mass and peak velocity (r $\approx0.64$) further reflects this connection. The positive correlation between ZAMS mass and both ejected mass and kinetic energy is expected since more massive stars build more massive cores and retain more gravitational binding energy. The inverse correlation with \rot, both for ejected mass (r $\approx-0.64$) and energy (r $\approx-0.49$), is due to angular momentum driven mixing enhancing winds and reducing envelop mass. These effects are consistent with theoretical expectations for rotating stars (e.g., \citealt{2012ARA&A..50..107L}; \citealt{2011A&A...530A.115B})


To ensure the robustness of the correlation matrix, we note that it is derived from the full set of approximately 40 models. The sample size, while moderate, is sufficient for exploratory analysis to identify statistically significant trends, particularly for strong correlations (e.g., r $\approx$ 0.95 for M$_{\text{eject}}$ and E$_{\text{KE,sh}}$). Weaker correlations (e.g., r $\approx$ 0.49) or those involving parameters like pulse duration may be less robust due to the limited sample size and potential influence of unmodeled factors. While correlation matrices are more commonly used in observational studies to identify patterns in data, they can also serve as a valuable diagnostic tool in theoretical investigations. In our case, we use the correlation matrix not as a substitute for physical modeling but to systematically highlight relationships between stellar and explosion properties across our multidimensional model grid. This approach allows us to identify which dependencies warrant deeper physical interpretation, particularly those that are non-trivial or nonlinear. Our focus remains on interpreting the strongest, physically meaningful trends using established principles of stellar evolution, rotation, and mass loss. This analysis suggests that the initial mass, rotation rate, and mass ejection strongly influence the energy dynamics and peak velocities of the explosion. In contrast, rotation tends to reduce the ejected mass and the energy output.

The time elapsed between two pulses ranges from a few days up to $\sim$ 2,000 years with a median value of $\sim$ 136 days.

In addition, we see a strong correlation between rotation rate and ejected shell composition: rapid rotation triggers enhanced rotationally-induced mixing that, in turn, leads to more chemically homogeneous evolution (CHE) and \cc- and \oo-rich envelopes. As a result, the ejected shells of rapidly rotating PPISN progenitors are \hh-poor, and, as such, their transients are eventually also expected to produce \hh-poor spectra akin to SN-I and SLSN-I in the most luminous cases. 

\begin{figure*}[h!t]
    \fig{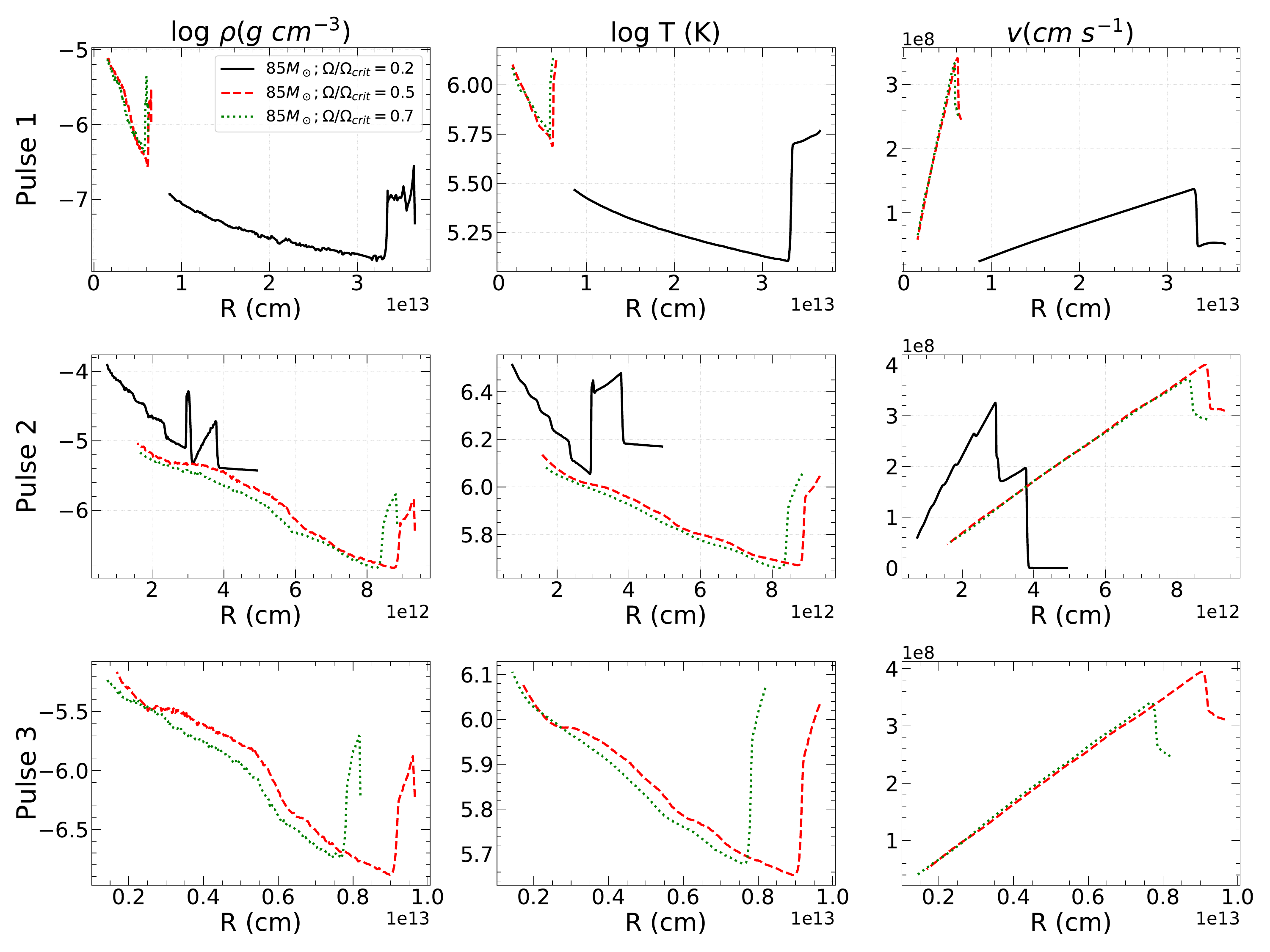}{\textwidth}{}
    \caption{Density (left), temperature (middle), and velocity (right) as a function of radius from the base (inner radius) to the outer edge of the unbound shell material in the pulses of the 85 \msun\ models with \rot $= 0.2$ (black solid lines), \rot $= 0.5$ (red dashed lines), and  \rot $= 0.7$ (green dotted lines). Profiles are plotted just after the onset of each pulse’s expansion phase, when velocities reach a maximum and homologous expansion begins.}
    \label{fig:pulses_85}
\end{figure*}

\begin{figure*}[h!b]
    \fig{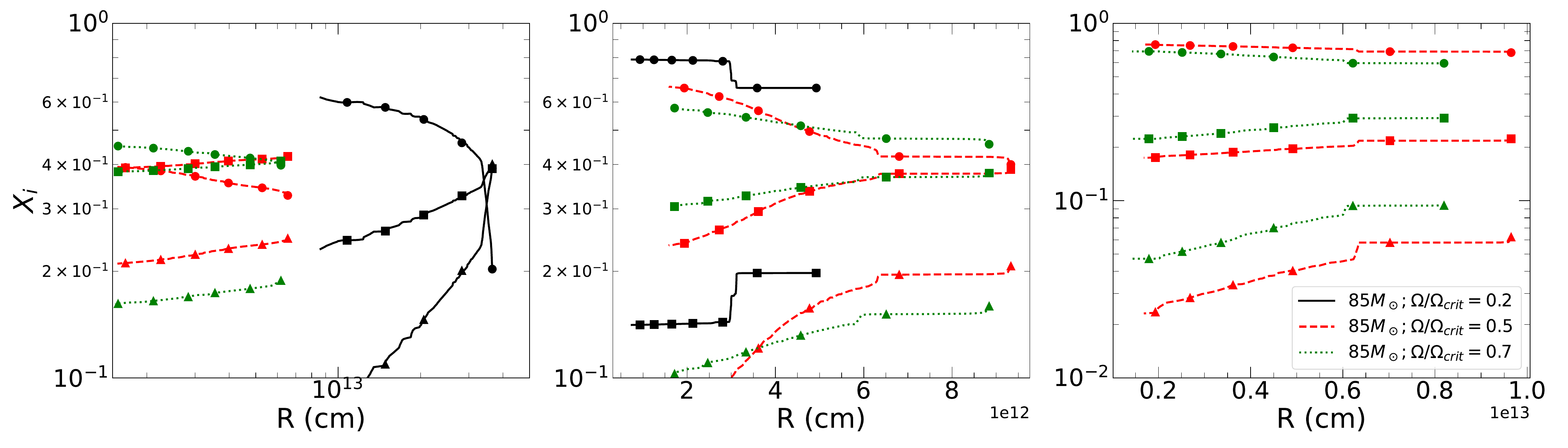}{\textwidth}{}
    \caption{The mass fraction of $^4\text{He}$ (triangle), $^{12}\text{C}$ (square), and $^{16}\text{O}$ (circle) as a function of radius from the center of the pulses of the 85 \msun\ models with \rot $= 0.2$ (black solid lines), \rot $= 0.5$ (red dashed lines), and  \rot $= 0.7$ (green dotted lines). Note that the mass fraction of $^4\text{He}$ for the  85 \msun\ model at \rot $= 0.2$ is $\approx 10^{-2}$, below the minimum mass fraction shown here.} \label{fig:abund_85}
\end{figure*}

\begin{figure*}[h!t]
\centering 
\fig{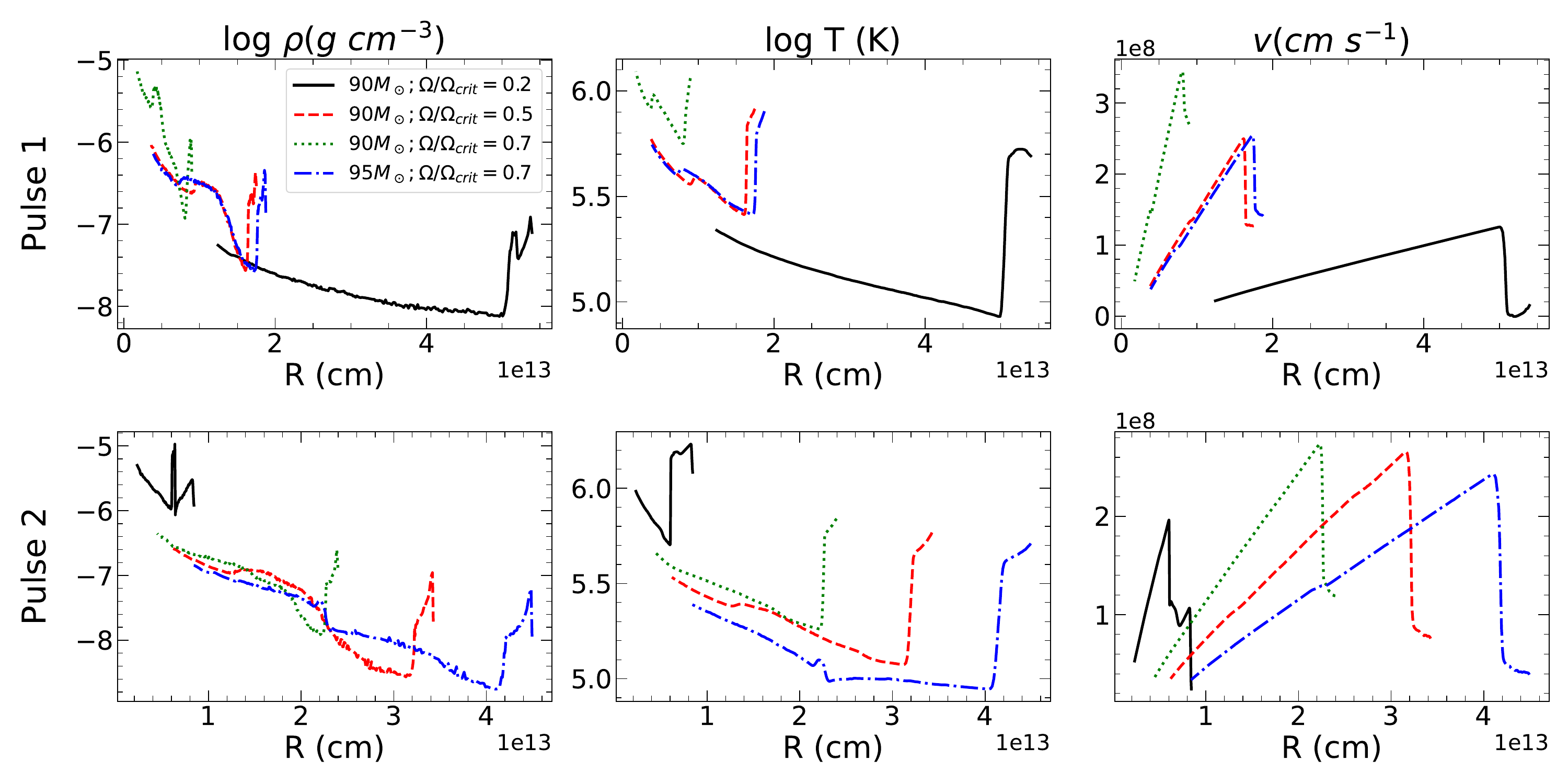}{\textwidth}{}
\caption{Density (left), temperature (middle), and velocity (right) as a function of radius from the base (inner radius) to the outer edge of the unbound shell material in the pulses of the 90 \msun\ models with \rot $= 0.2$ (black solid lines), \rot $= 0.5$ (red dashed lines) and  \rot $= 0.7$ (green dotted lines) and the 95 \msun\ model with \rot $= 0.7$ (blue dash-dot lines). Profiles are plotted just after the onset of each pulse’s expansion phase, when velocities reach a maximum and homologous expansion begins.} \label{fig:pulses_90M}
\end{figure*}

\begin{figure*}[h!t]
\centering 
\fig{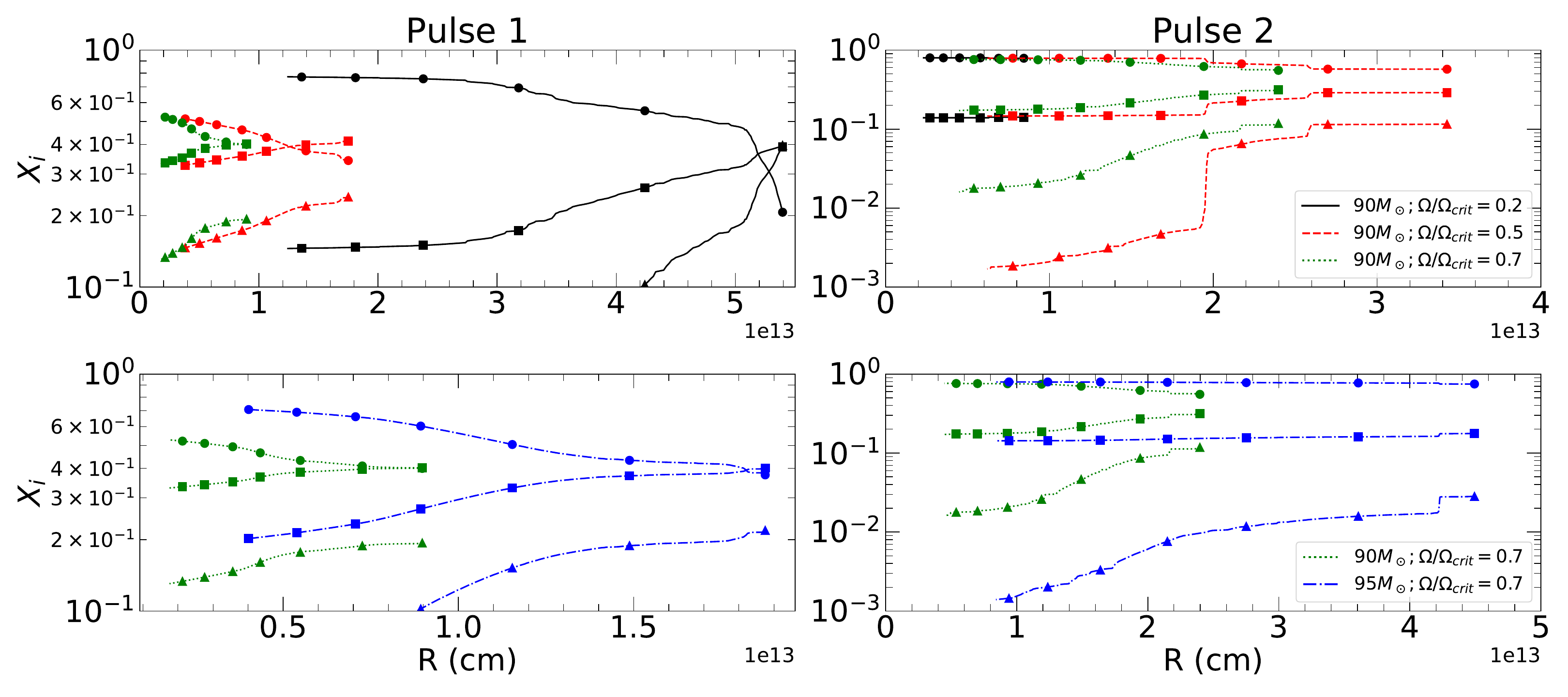}{\textwidth}{}
\caption{The mass fraction of $^4\text{He}$ (triangle), $^{12}\text{C}$ (square), and $^{16}\text{O}$ (circle) as a function of radius from the center of the two pulses. (Top row) Model 90 \msun\ with various rotational velocities: \rot $= 0.2$ (black solid lines), \rot $= 0.5$ (red dashed lines) and  \rot $= 0.7$ (green dotted lines). (Bottom row) With the same rotation rate at \rot $= 0.7$, Model 90 \msun\ and 95 \msun\ are displayed in green dotted and blue dash-dot lines, respectively.} \label{fig:abund_90}
\end{figure*}

\begin{figure*}[h!t]
    \fig{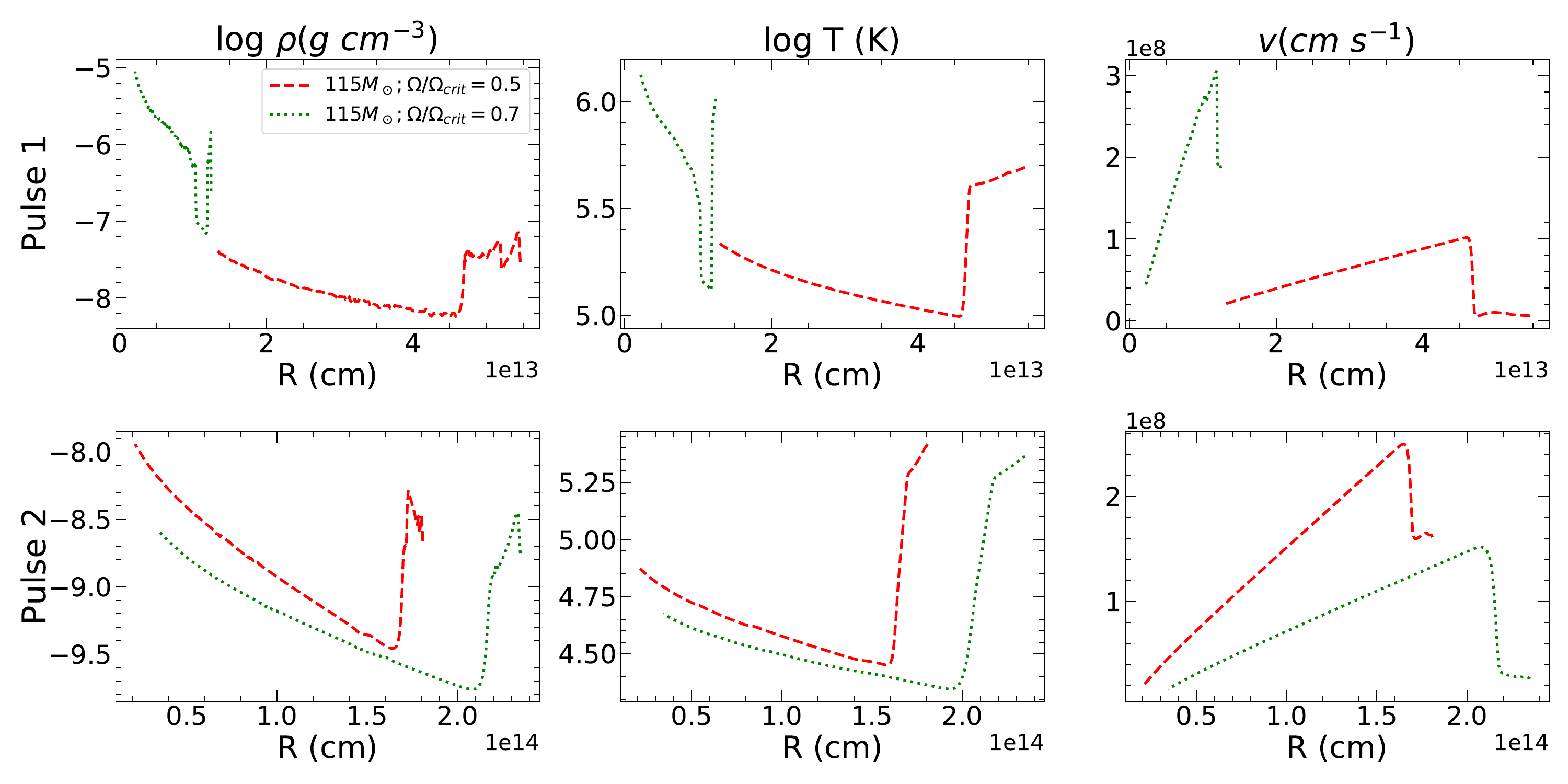}{\textwidth}{}
    \caption{Density (left), temperature (middle), and velocity (right) as a function of radius from the base (inner radius) to the outer edge of the unbound shell material in the pulses of the  115 \msun\ models with \rot $= 0.5$ (red dashed lines) and  \rot $= 0.7$ (green dotted lines). Profiles are plotted just after the onset of each pulse’s expansion phase, when velocities reach a maximum and homologous expansion begins.} \label{fig:pulses_115M}
\end{figure*}

\begin{figure*}[h!t]
    \centering 
    \includegraphics[width=0.8\textwidth]{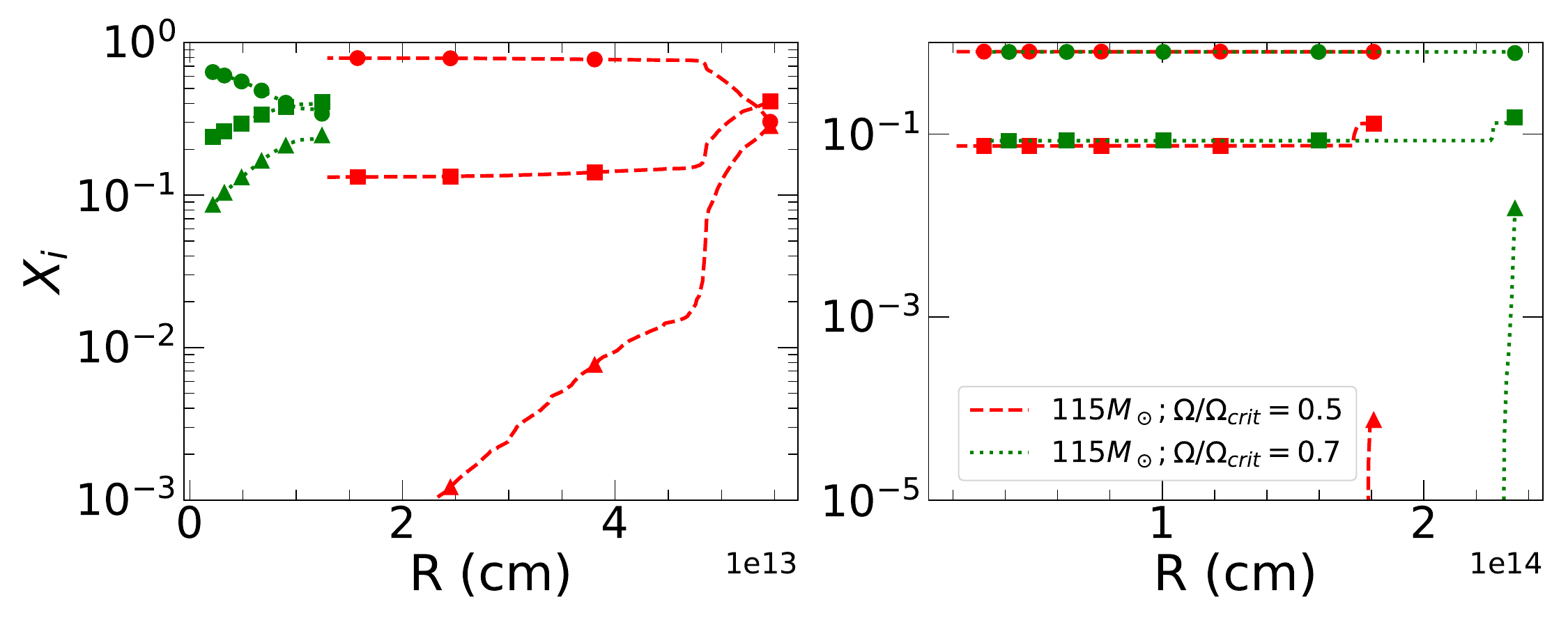}
    \caption{The mass fraction of $^4\text{He}$ (triangle), $^{12}\text{C}$ (square), and $^{16}\text{O}$ (circle) as a function of radius from the center of the two pulses of the 115 \msun\ models with \rot $= 0.5$ (red dashed lines) and  \rot $= 0.7$ (green dotted lines).} \label{fig:pulses_115M_abund}
\end{figure*}

In order to compare the effects of rotation on the dynamics and composition of the PPISNe progenitors and their ejected materials, we present from Figures \ref{fig:pulses_85} through \ref{fig:pulses_115M_abund} some details on the pulse profiles of the models 85, 90, 95, and 115 \msun\ with different rotational rates and the composition of their ejected materials, accordingly. The pulse profiles (e.g., density, temperature, and velocity) plotted in these figures represent the moment just after the onset of each pulse expansion's phase, when velocities reach a maximum and homologous expansion begins, which captures the initial condition of the ejecta. To illustrate the rotational effects on PPISNe at fixed M$_{\text{ZAMS}}$ as well as M$_{\text{ZAMS}}$ effects at fixed rotation, we pick the models in Figures \ref{fig:pulses_90M} and \ref{fig:abund_90} for further discussion. Figure \ref{fig:abund_90} shows the mass fractions of \he, \cc, and \oo\ as functions of radius for two pulses. The top panels compare three 90 \msun\ models at \rot $= 0.2,\ 0.5,$ and 0.7. The bottom panel compares two different ZAMS masses 90 \msun\ and 95 \msun\ at fixed rotation rate (\rot $= 0.7$). At \rot $= 0.5$, the outer ejecta is \he-rich, while \cc\ and \oo\ are abundant throughout. High rotation also promotes homologous shell compositions in later pulses, as seen in the bottom right panel of Figure \ref{fig:abund_90}. Additionally, the first ejection travels faster in high-rotation models, causing subsequent pulses to overtake earlier material. 

For the rotating 115 \msun\ models shown in Figures \ref{fig:pulses_115M} and \ref{fig:pulses_115M_abund}, the outer radius of the second pulse is larger than that of the first pulse, unlike the other models (Figures \ref{fig:pulses_85}-\ref{fig:abund_90}). This is likely because the star expanded after the first pulse, leading to an increased radius after the first ejection and resulting in a larger outer radius at which the second ejection started.

The density profiles exhibit a typical shocked ejecta structure, characterized by a sharp peak and the concentration of most ejected mass in a narrow, dense shell expanding at SN-like velocities (a few \kmsone in most cases). The velocity profiles indicate homologous expansion (u $\propto$ r) for the shell material internal to the location of peak velocity. At the time of ejection, during the early expansion phase, the peak shell temperatures range from 100,000 to a few million \KK, suggesting the material is extremely hot and highly ionized. These temperatures are expected to drop rapidly during the ``photospheric" phase, a few days to months after each pulsation, primarily due to adiabatic and radiative cooling processes. This cooling and density reduction will drive the recombination of elements such as \he, \cc, \nn, and \oo, producing spectral lines with P Cygni profiles as the shell expands. 

Additionally, we select three PPISN models as strong candidates for shell collisions. These models exhibit later pulses that eject material at a higher speed and with greater kinetic energy than the preceding pulses. Another key criterion is that the model has pulses that occur relatively soon after each other to ensure that they are still at relatively high density when the shells collide, thereby leading to strong interaction and potentially luminous transients. Based on the details in Table \ref{tab:shells}, the selected candidates are M$_{\text{ZAMS}}$ = 85\msun, \rot $= 0.5$, M$_{\text{ZAMS}}$ = 85 \msun, \rot $= 0.7$, and M$_{\text{ZAMS}}$ = 95\msun, \rot $= 0.5$. All three models involve the collision of the second pulse with the first. 

In Section \ref{sec:lightcurve}, we use radiation transport simulations to model the radiative properties of rotating PPISN shell ejections and explore potential similarities with observed transients.

\subsection{Comparison with \cite{2017ApJ...836..244W}}


In this subsection, the primary goal of the comparisons is to interpret the effects of rotation and validate our model outputs in the context of previously published non-rotating work. The \cite{2017ApJ...836..244W} models are not used in our simulations or analysis directly; rather, the purpose of citing the study is to compare their extensive set of non-rotating PPISN models with our rotating counterparts. \cite{2017ApJ...836..244W}'s models were computed using the KEPLER stellar evolution code (\citealt{1978ApJ...225.1021W}; \citealt{2007PhR...442..269W}), while ours are calculated with \texttt{MESA}. Here, we selected several Woosley models with similar \he-core masses prior to the onset of PPI to compare with our rotating models and assess how rotation alters the dynamics, enegetics, and composition of ejected shells. 

Tables \ref{tab:lowZ} and \ref{tab:shells} include data from Woosley’s work, specifically three non-rotating PPISN models at 10\% \zsun. These models — T110B with $1/4$ standard mass loss rate, T123A with $1/2$ standard mass loss rate, and T130D with no mass loss — were selected due to their comparable \he-core masses: 52 \msun\ for our 85 \msun\ model at \rot $= 0.7$, 55 \msun\ for our 90 \msun\ model at \rot $= 0.5$ and $0.7$, and 60 \msun\ for our 90 \msun\ model at \rot $= 0.2$, respectively. Note that the stronger mass ejections in all Woosley's non-rotating models, compared to \texttt{MESA} rotating counterparts, stem from a combination of different mass loss assumptions (the standard mass loss multiplied by factors of 1, 1/2, 1/4, 1/8, and 0) and the absence of angular momentum effects. In contrast, our rotating models experience rotationally enhanced winds and chemically homogeneous evolution, reducing the retained envelope mass.

Model T110B, a non-rotating PPISN with a ZAMS mass of 110 \msun\ and a standard mass loss factor of 1/4, shares similar \he-core mass with our 85 \msun\ model at high rotation (\rot $= 0.7$). T110B ejects a substantial 35.3 \msun\ in its first pulse at an average velocity of $\sim1300$ \kmsone, followed by a two-year quiescent period, producing \hh-rich ejecta due to its retained envelope. In contrast, the rotating model, undergoing three pulses, ejects less mass ($\sim$1.24 \msun) in its second of the three pulses, with higher velocities ($\sim 3600$ \kmsone) and shorter quiescent periods (7 days after the first pulse and $\sim$ 20 days after the second). Its first pulse, lasting $\sim$2 days, ejects a \he-rich envelope, while later pulses are \oo-rich due to a large \co\ core, driven by rotationally induced mixing (e.g., meridional circulation). Rotation reduces envelope mass via centrifugal-driven mass loss and accelerates core dynamics, shortening quiescent times, while T110B's higher ejecta mass reflects its larger ZAMS mass and mass loss assumptions.

\begin{figure*}[h!btp]
    \centering 
    \includegraphics[width=0.7\textwidth]{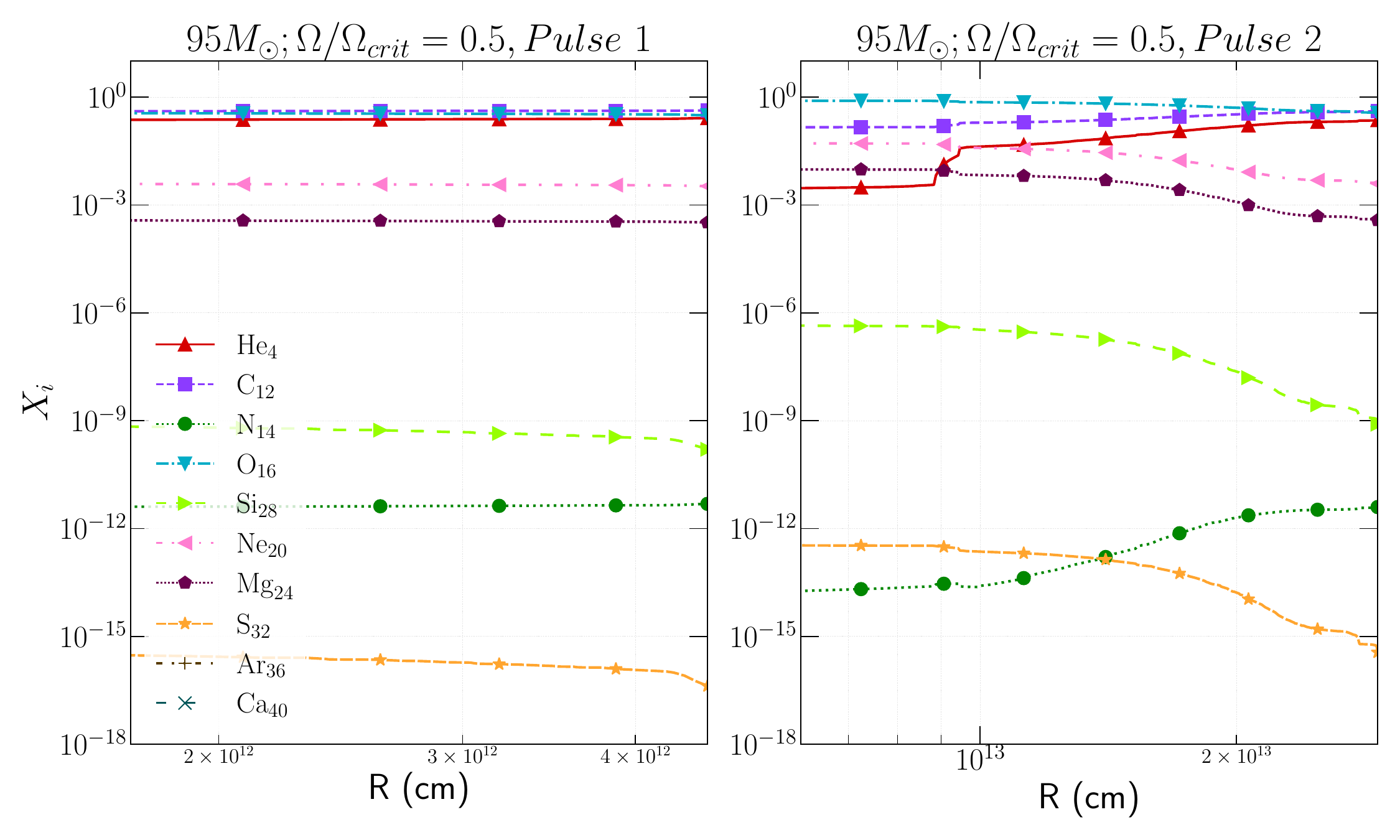}
    \caption{Logarithmic plots showing the mass fraction (X$_i$) of various isotopes ($^{4}\text{He}, ^{12}\text{C}, ^{14}\text{N}, ^{16}\text{O}, ^{28}\text{Si}, ^{20}\text{Ne}, ^{24}\text{Mg}, ^{32}\text{S}, ^{36}\text{Ar}, ^{40}\text{Ca}$) as a function of radius (cm) in the 95~M$_{\odot}$, \rot $= 0.5$, at two stages: Pulse 1 (left) and Pulse 2 (right). \label{fig:combined_abund}}
\end{figure*}

Next, we examine Model T123A, a non-rotating PPISN with a ZAMS mass of 123 \msun\ and a mass loss factor of 1/2, sharing a comparable \he-core mass ($\sim$ 55 \msun) with our 90 \msun\ rotating models (\rot $=0.5$ and $0.7$). All three undergo two pulses of mass ejections, with the second pulse ejecting the most mass but at relatively lower kinetic energy. However, T123A ejects more mass overall (e.g., $\sim$ 17.8 \msun\ of \hh-rich envelope in the first pulse and 6.4 \msun\ of \he-rich materials in the second pulse). Its first pulse lasts for $\sim$ 120 days, followed by a century-long quiescent period before the second pulse and collapse into a black hole. In contrast, the rotating models eject less mass ($\sim$ 1--2 \msun\ per pulse) with shorter quiescent intervals ($\sim$ 40--205 days) and higher peak velocities ($\sim$ 3000--4000 \kmsone\ versus T123A's $\sim$ 1000 \kmsone). Their \he- and \oo-rich ejecta result from rotation-driven mixing, while T123A's first pulse is H-rich. Rotation reduces envelope mass via centrifugal-driven mass loss and shortens quiescent times through mixing (e.g., meridional circulation), while T123A's higher ejecta mass results from its larger ZAMS mass and lower mass loss rate.

Finally, we compare Model T130D, featuring no rotation and no mass loss, having \he- and \co-core masses ($\sim$ 60 \msun\ and $\sim$ 54 \msun) similar to our 90 \msun\ rotating model (\rot $= 0.2$). T130D undergoes three powerful pulses, with the first ejecting 70 \msun\ of H-rich envelope at $1.5 \times 10^{51}$ erg, resulting in classic SN IIP-like light curves without radioactive heating, which is present in normal SN IIP events emanating from red supergiant progenitors. After $\sim$ 3,300 years, the second pulse ejects 7.7 \msun\ of \he, \cc, and \oo\ at $\sim 10^{51}$ erg. Eight months later, a third pulse ejects 13.5 \msun\ (mostly \he/\cc/\oo), colliding with the second pulse's ejecta at $\sim 10^{15}$ cm, with velocities $\sim$ 1000-2000 \kmsone\ across pulses. In contrast, the rotating model ejects $\sim$ 10 \msun\ total, with the first pulse (lasting 5 days) expelling 7 \msun\ of \he-rich materials from $\sim 5 \times 10^{13}$ cm, and the second pulse (after 1940 years) ejecting $\sim$3 \msun\ of material of \oo-rich (80\%) and \cc-rich material with no \he, at a peak velocity of 1980 \kmsone\ for less than a day. This second pulse, with $\sim 1 \times 10^{50}$ erg, collides with earlier shells, potentially resembling \hh-poor Type I SNe (Ib/Ic or Ibn/Icn). Rotation lowers ejecta mass by stripping the envelope and produces \oo-rich ejecta via mixing.

\section{Light Curves of \hh-poor PPISN shells} \label{sec:lightcurve}

In this section, we present the light curve results for the two \hh-poor PPISN shells and their collision in one of our PPISN models. The model was selected based on the following criteria: (1) the star undergoes at least two mass ejections driven by the pair instability, (2) the second ejection has higher energy and velocity than the first, and (3) the time interval between the two ejections ($\Delta \text{t}_{\text{pulse}}$) is short. These conditions ensure that the second shell quickly overtakes and interacts with the first in a dense environment, facilitating the development of strong shocks and potentially producing a luminous transient. Based on these criteria, we selected a 95 \msun\ model with \rot $= 0.5$ and a metallicity of 10\% \zsun for post-processing. In this model, the second ejection is $8 \times 10^{49}$ erg more energetic and reaches a slightly higher peak velocity than the first pulse, with a time separation of 26 days between the two pulses.

\begin{figure}[h!]
\centering 
\includegraphics[width=0.5\textwidth]{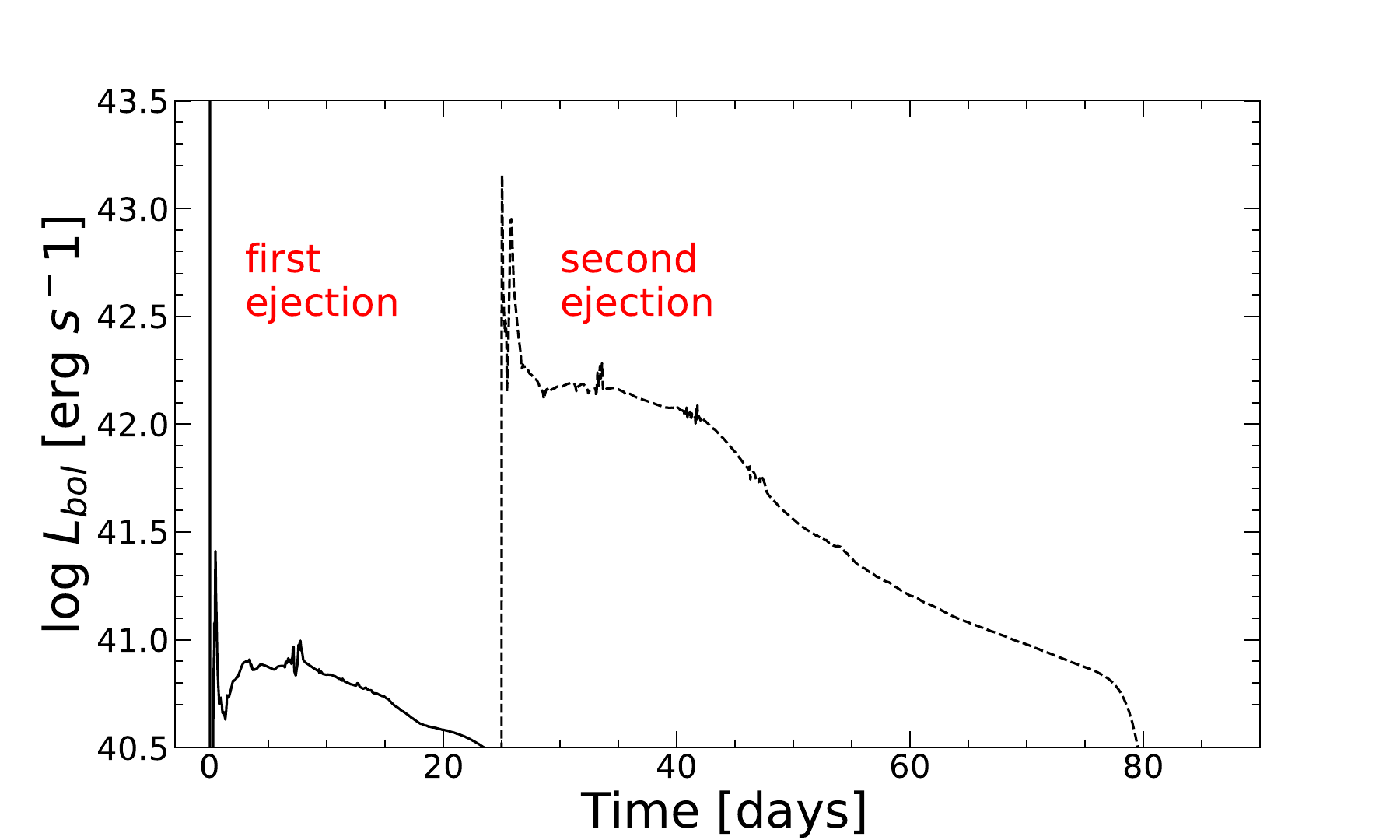} 
\caption{ The bolometric light curve of the two shell ejections for the 95 \msun\ model with \rot $= 0.5$ as a function of time in days since the ejection of the first shell (solid). 26 days later, the second shell ejection (dashed) occurs and remains bright above $\sim 3\times 10^{40}$ erg s$^{-1}$ for the next 50 days. \label{fig:pulse_lc}}
\end{figure}

In a post-processing step, we extract the hydrodynamic profiles of the unbound mass from the first and second ejections of the rotating 95 \msun\ model produced with \texttt{MESA}, and use them as inputs for \texttt{STELLA}. \texttt{STELLA} is first used to calculate the light curve of the first pulse and to model its evolution up to 100 days. We then simulate the second pulse (again with \texttt{STELLA}), propagating it until its expanding radius reaches the point where it is expected to interact with the previously ejected shell, assuming homologous expansion of the first ejecta. Figure~\ref{fig:combined_abund} shows the composition of the two ejecta shells. In this model, after helium is depleted in the core, the star retains a mass of 58 \msun. The first pair-instability event occurs when the star reaches a maximum central temperature of $2.8 \times 10^{9}$ \KK and a central density of $1.5 \times 10^{6}$ \gcmthree. The star undergoes a strong pulse, ejecting 0.78 \msun\ of material composed of 24\% \he, 41\% \cc, and 34\% \oo, with a total energy of $1.8 \times 10^{49}$ erg. This ejecta reaches a peak velocity of approximately 2000 \kmsone\ and is released over a timescale of less than one day. About 26 days later, the star experiences a second, more powerful pulse, with an energy of $7.8 \times 10^{49}$ erg and a peak velocity of approximately 2100 \kmsone, lasting about 1.5 days. This second ejection expels nearly 4 \msun\ of material, predominantly composed of 62\% oxygen. Both ejections are completely devoid of hydrogen. The evolution of the bolometric light curves for the two ejections is shown in Figure~\ref{fig:pulse_lc}, illustrating the 26-day interval between them.

\begin{figure}[htbp]
\centering 
    \includegraphics[width=0.5\textwidth]{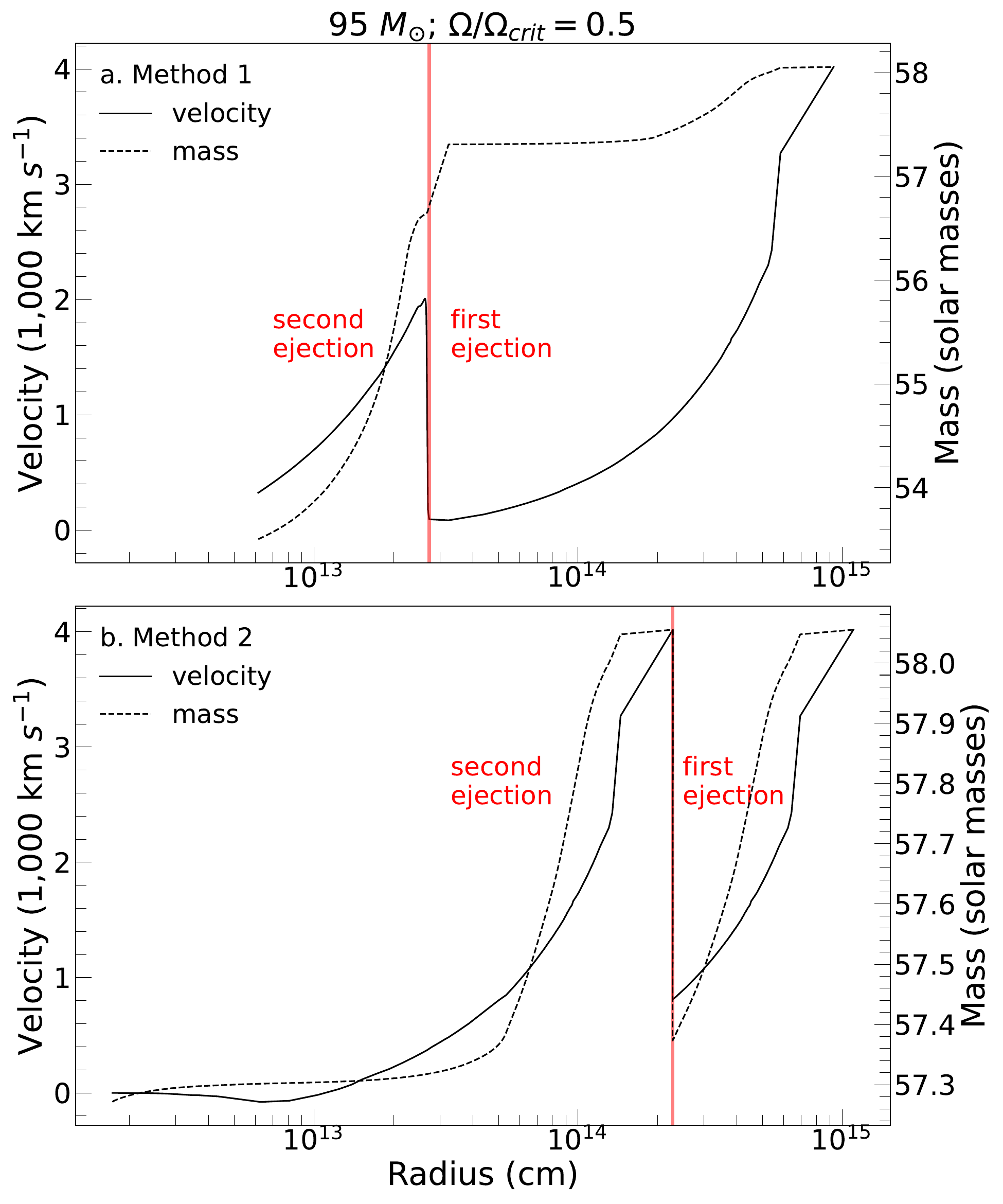} 
    \caption{Velocity and mass structure of the two-shell collision for the 95 \msun\ model with \rot $= 0.5$. 
\textbf{(a)} Top panel: Following Method 1, the red line indicates the discontinuity at R$ = 10^{13}$~cm where the second ejecta catches up to the tail of the first. The mass axis shows the total enclosed mass of $\sim 58.5$ \msun, equivalent to the mass remaining after the \he-depletion phase. 
\textbf{(b)} Bottom panel: Following Method 2, on day 31, the two ejecta have reached a radius of $10^{14}$~cm, where the faster-moving second ejecta collides with the first. 
\label{fig:combined_velocity}}
\end{figure}

\begin{figure}[h!b]
    \centering 
    \includegraphics[width=0.5\textwidth]{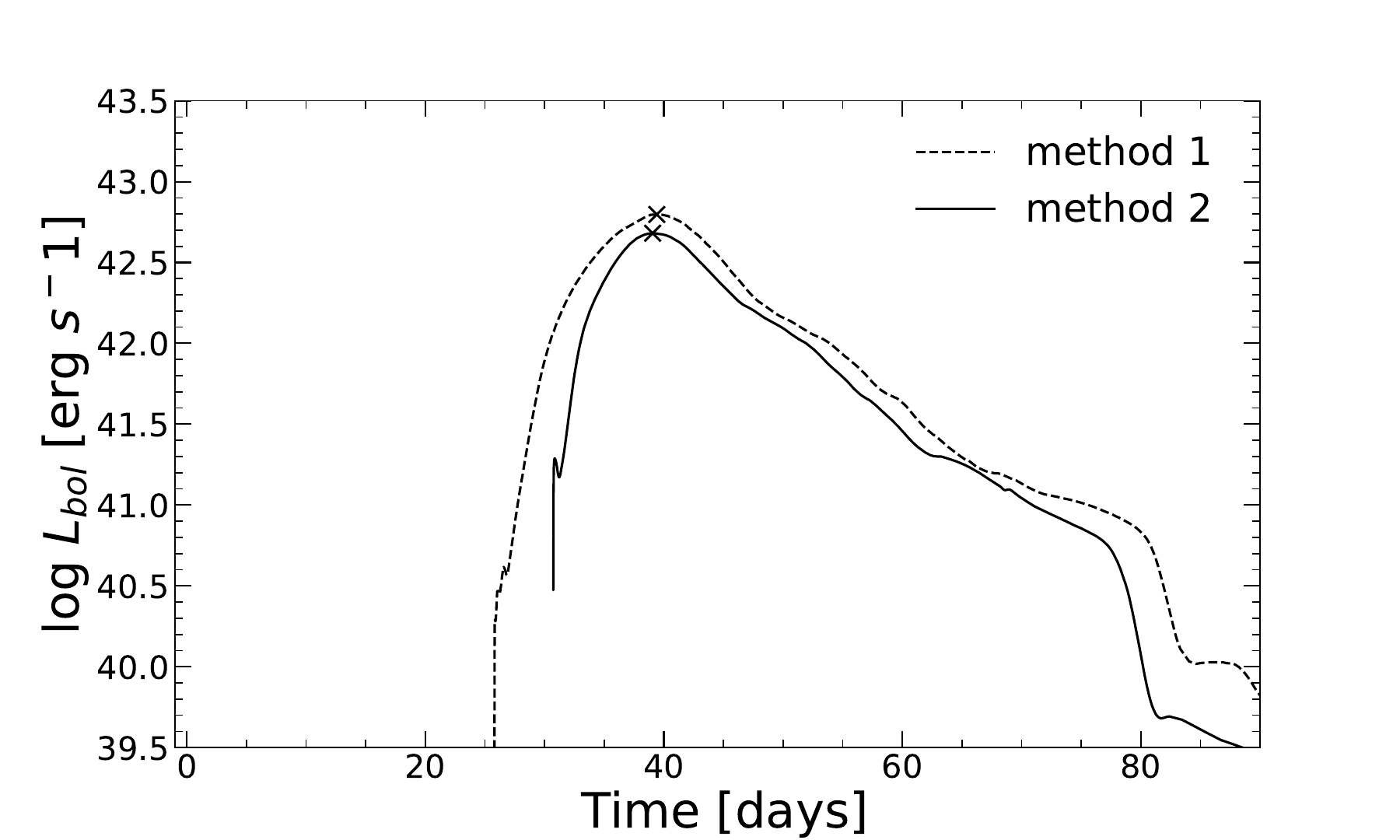}
    \caption{The bolometric light curve of the two-shell collision for the rotating $95~$ \msun model at \rot $=0.5$ using two methods mentioned where the time axis starts from the first ejection. The 'x' symbols label the peak luminosity. \label{fig:combined_lc}}
\end{figure}

The interaction time of the two \hh-poor shells is defined as t$_{\text{int}} = \Delta \text{t}_{\text{pulse}} + \text{dt},$ the time difference between two pulses plus the time it takes for the second pulse to catch up to the first. We follow two methods to determine t$_{\text{int}}$ and model the light-curve evolution of the colliding shells. For the first method (denoted Method 1), we combine the hydrodynamic profiles of pulse one on day 26 (the time difference between two ejections) and pulse two at the time of ejection (day 1). This introduces the profile of pulse 2 in the inner boundary while that of pulse 1 is still in the grid. The velocity structure of the two ejections for this method is shown in Panel a), Figure \ref{fig:combined_velocity}. For the second method (denoted as Method 2), we select the profile when the tip of pulse two on day 6 catches up with the densest region of pulse one on day 31 at a radius of R $=2.3 \times 10^{14}$ cm. In Panel b) of Figure \ref{fig:combined_velocity}, the velocity discontinuity at $10^{14}$ cm shows where this fast-moving ejecta of the second explosion begins to collide with the moving ejecta of the first. At that position, the velocity of the second outburst is approximately 8 times that of the first one. The time shift between Methods 1 and 2 is approximately 5 days. The time it takes for the second pulse to catch up with the first is approximately 14 days as its peak encounters the steep density gradient of the first pulse. However, the interaction practically begins earlier, as the dense part of the second starts shocking the inner, rising density gradient of the first pulse. The light curves of the collision for the two methods are presented in Figure \ref{fig:combined_lc} relative to the time of the model's first ejection. The collision between the two shells produces an intermediate-luminosity light curve that is consistent with classical SN levels but with a bright duration that lasts $\sim$50 days. The bolometric luminosity of the collision peaks at $\sim 4.8 - 6.3 \times 10^{42}$ erg s$^{-1}$. The two methods produce similar light curves, where the first method is slightly more luminous.

Regarding the timing of subsequent pulses, the 95 \msun\ model with \rot$=0.5$ undergoes two major mass ejections. The interaction of two shells, producing a third signal, occurs approximately 14 days after the second ejection. The onset of the core collapse occurs more than a century after the second ejection, based on \texttt{MESA} simulation. Thus, the third optical signal appears well before the onset of core collapse.

\begin{figure*}[h!t]
    \centering 
    \includegraphics[width=1\textwidth]{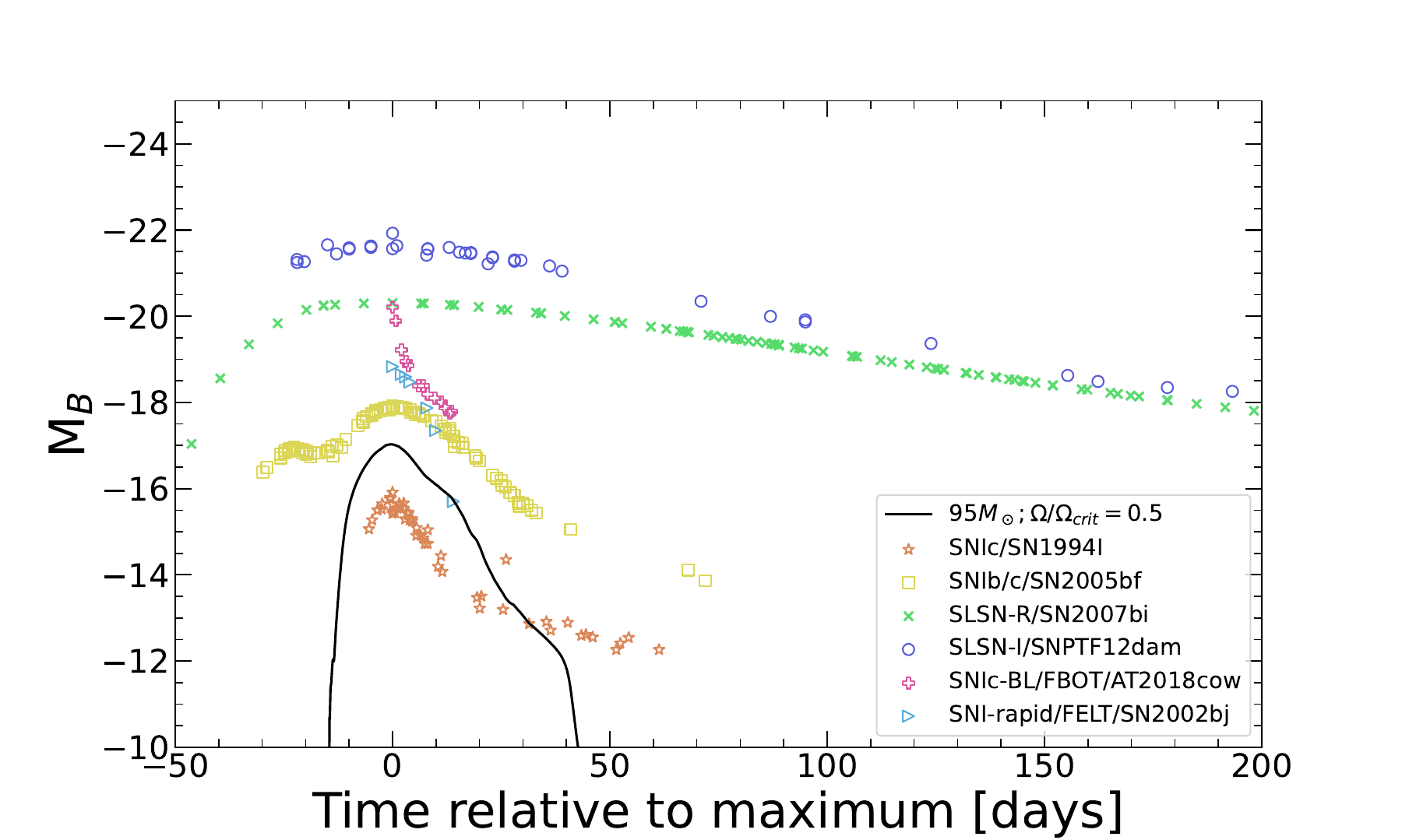}
    \caption{Comparison of the B-band absolute magnitude evolution with other observed supernovae and the shell collision model for the $95~$ \msun at \rot $=0.5$. The phase is with respect to the maximum brightness. The adopted maximum light dates in MJD and Distance (Mpc) for each supernova are, respectively: SN1994I: 49449.9, 7.9 (\citealt{1994ApJ...437L.115I}, \citealt{1996AJ....111..327R}); SN2005bf: 53497.0, 83.3 (\citealt{2018A&A...609A.135S}); SN2007bi: 54145.3, 548.9 (\citealt{2010A&A...512A..70Y}, \citealt{2012PASP..124..668Y}); SNPTF12dam: 56090.2, 462.8 (\citealt{2017ApJ...835...64G}); SN2018cow: 58286.9, 62.4 (\citealt{2018ApJ...865L...3P}); SN2002bj: 52333, 53 (\citealt{2010Sci...327...58P}).\textit{Note that the phase reported for SN2005bj is relative to the discovery and not the maximum.} \label{fig:olc_slc}}
\end{figure*}

In terms of visualizing the overall optical pattern, combining the light curves of the two ejections and their interaction on a single time axis is currently limited by computational constraints in post-processing hydrodynamics profiles with \texttt{STELLA}. Qualitatively, we expect the first pulse to produce a short-lived optical signal due to its low ejected mass (0.78\msun) and energy ($\sim 1.8 \times 10^{49}$ erg), peaking at $\sim 10^{41} \text{ erg s}^{-1}$. Around 26 days later, the second pulse, with a higher mass (4 \msun) and energy ($\sim 7.8 \times 10^{49}$ erg), produces a brighter signal, peaking at $\sim 1.3 \times 10^{42} \text{ erg s}^{-1}$. The second pulse then catches up to the first pulse within $\sim14$ days, leading to a shell-shell interaction that generates a more luminous transient, peaking at $\sim 4 \times 10^{42} \text{ erg s}^{-1}$. Speculatively, this sequence in our 95 \msun\ model at 50\% critical rotation, exhibiting three optical signals of pulse ejections, could explain SN impostors or multi-peak events like SN 2009ip and SN 2016bdu. Both of these events have been suggested as arising from non-terminal eruptions and subsequent shell-to-shell collisions driven by PPI, producing high velocity ejecta and radiative outbursts from kinetic energy dissipation \citep{2013ApJ...767....1P, 2018MNRAS.474..197P}.

\begin{figure*}[h!t]
    \centering 
    \includegraphics[width=1\textwidth]{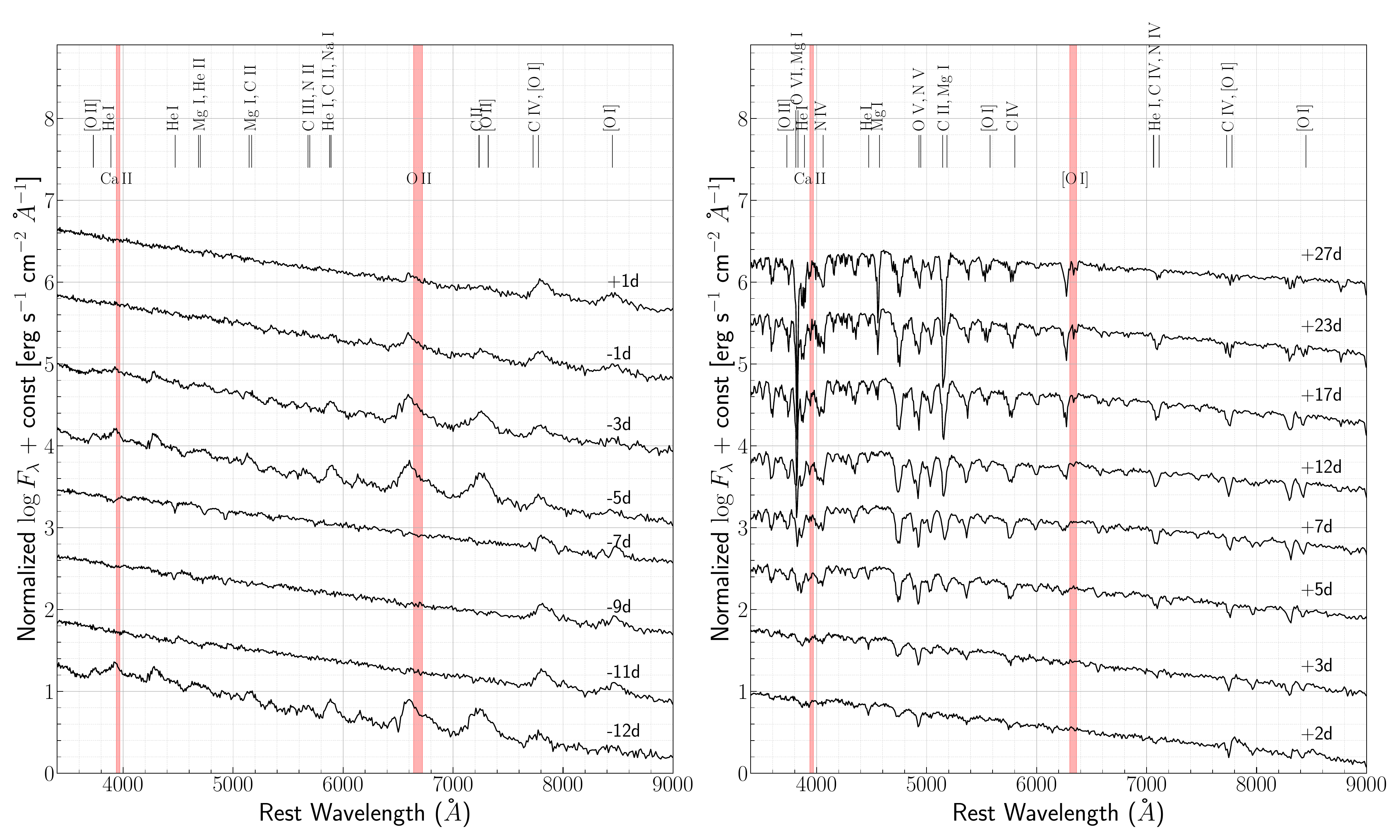}
    \caption{Spectroscopic evolution of the \hh-poor shell collision for the 95 \msun\ model rotating at \rot $=0.5$ produced by the \texttt{SuperLite} code between -12 days and +27 days of the peak luminosity of the shell--collision light curve. The red bands represent key features: the \caII\ band represents the calcium lines at $\mathrm{\lambda 3939}$ and $\mathrm{\lambda 3969 \text{\AA}}$; the \oII\ band represents the oxygen multiplet at $\mathrm{\lambda 6641}$ and $\mathrm{\lambda 6721 \text{\AA}}$; and the \ooI\ band represents the oxygen forbidden lines at $\mathrm{\lambda 6300}$ and $\mathrm{\lambda 6363 \text{\AA}}$.}\label{fig:spec}
\end{figure*}

In Figure~\ref{fig:olc_slc}, we compare the absolute B-band magnitudes of our model to those of observed Type Ib/c supernovae, SLSNe, and two fast-evolving luminous transients (FELT and FBOT). Since both light curve extraction methods produced consistent results, we use Method 1 for the light curve shown in this figure. The model’s peak brightness (M$_{\text{B}} \sim -17$) is fainter than that of SN PTF12dam, SN 2007bi, AT2018cow, and SN 2002bj (M$_{\text{B}} \sim -18$ to $-22$), but comparable to SN 1994I and SN 2005bf (M$_{\text{B}} \sim -16$ to $-18$). Among the comparison sample, SN PTF12dam stands out as one of the brightest, consistent with it being a \hh-poor SLSN-I event. Regarding the light curve shape, our model exhibits a rapid rise to peak, similar to that of SN 1994I, followed by a steep post-peak decline comparable to both SN 1994I and SN 2005bf. While SN 2005bf begins to fade rapidly around t $\sim 50$ days \citep{2007ApJ...666.1069M}, eventually becoming very faint at later phases, our model declines even more quickly, reaching M$_{\text{B}} \sim -10$ less than 50 days after peak. In contrast, SLSN-I PTF12dam and SLSN-R SN 2007bi display relatively flat, long-lasting light curves, with PTF12dam maintaining high luminosity for an extended period. We also compare the model spectra to the two special transients, AT2018cow (i.e. FBOT) and SN 2002bj (i.e. FELT). AT2018cow, with a peak luminosity of M = -21.8 mag and a rapid rise of over 5 mag in 3.3 days, is an unusual transient, while SN 2002bj, one of the earliest fast transient identified, was once postulated as a Type ".Ia" event but its peak intrinsic brightness exceeding -18 mag suggests otherwise. Both transients show rapid rise and steep decline, and while our model's light curve, peaking at 14 days after the collision, shares some similarities, the luminous transients are far brighter. Based on these comparisons, the light curve produced by our PPISN shell collision model more closely resembles that of typical Type Ib/Ic supernovae than SLSNe or fast-evolving luminous transients.

\section{Spectroscopic Evolution of \hh-Poor Shell Collision} \label{sec:spec}

Figure~\ref{fig:spec} shows the spectroscopic evolution of the shell collision for the rotating 95 \msun\ model, spanning from 12 days before peak luminosity ($-12$ days) to 27 days after peak ($+27$ days), as calculated using the \texttt{SuperLite} code. We also use Method 1 for the spectra shown in this section when the luminosity peaks at day 13. The synthetic spectra are plotted over the wavelength range 3000–9000$\text{\AA}$, covering the visible portion of the electromagnetic spectrum.
Before peak luminosity, the spectra are characterized by broad \cII\ features blending with \oII\  emission lines at $\lambda 7231\text{\AA}$ and $\lambda 7320\text{\AA}$, respectively. Forbidden \ooII\ features are also present at $\lambda 7776\text{\AA}$ and $\lambda 8448\text{\AA}$.
At phases $-12$ days, $-5$ days, $-3$ days, and $-1$ days, we detect a mixture of \heI, \cII, and \naI\ lines around $\lambda 5876\text{\AA}$, which weaken over time. A broad, asymmetric feature with a P-Cygni profile appears around $\lambda 6600\text{\AA}$ in the early spectra, except between $-11$ days and $-7$ days, where instead two \oII\ lines at $\lambda 6641\text{\AA}$ and $\lambda 6721\text{\AA}$ are observed toward the red wing.
Weak \cII\ and \mgI\ emission lines at $\lambda 5145\text{\AA}$ and $\lambda 5167\text{\AA}$, respectively, become featureless approximately one day before peak.
At later phases, strong absorption features dominate the spectra, consistent with the cooling of the SN ejecta. Significant absorption lines include a blend of \oIV\ and \mgI\ around $\lambda 3811\text{\AA}$, \mgI\ at $\lambda 4571\text{\AA}$, \oV\ and $\text{N\,\textsc{v}}$ near $\lambda 4930\text{\AA}$, and \cII\ and \mgII\ around $\lambda 5145\text{\AA}$. A transition from emission to absorption is also seen in the forbidden \ooII\ line at approximately $\lambda 8448\text{\AA}$.

This spectral evolution reflects the underlying physical conditions of the shell collision. Before peak luminosity, the spectra are dominated by emission lines from moderately ionized species (e.g., \cII, \oII), produced by the interaction-driven heating of the outer ejecta. As the shock heats the material, it excites and ionizes the surrounding gas, leading to the observed emission features and P-Cygni profiles. As the system expands and cools after peak, the ionization state drops, and absorption features from neutral and singly ionized elements (e.g., \mgI, \oI, \cII) become more prominent. This transition from emission to absorption, and the overall spectral cooling, is consistent with the decline of the bolometric light curve after the collision-driven luminosity peak.


To compare the \texttt{SuperLite} spectra with observations, we use a Python-based program, \texttt{Duperfit} (\citealt{duperfit}), a spectral classification tool based on the \texttt{Superfit} algorithm \citep{2005ApJ...634.1190H}. \texttt{Duperfit} interprets an observed spectrum as a combination of supernova and host galaxy light and fits the input spectrum against a library of archival SN templates according to the following form:

\begin{equation}
    o_{\text{mod}}(\lambda; z) = Ct(\lambda; z) 10^{-0.4A_V r(\lambda; R_V)}+ Dg(\lambda; z)
\end{equation}

where $A_V$ (the total extinction in the V-band), $R_V$ (extinction parameter, 3.1 by default), and $A(\lambda; R_V)$ (wavelength-dependent extinction) adjust for relative reddening $r(\lambda; R_V) \equiv A(\lambda; R_V)/A_V$. Additionally, $t(\lambda; R_V)$ (some archival template) and $g(\lambda; R_V)$ (host galaxy) share the same redshift, while C and D scale the SN and host galaxy components. The program allows the optimization of the redshift via a grid search. \texttt{Duperfit} uses a least-squares method for fitting, with optional weighting from inverse variance, telluric-excluded spectra, or user-defined weight files. 

\begin{figure*}[h!t]
    \centering 
    \includegraphics[width=1\textwidth]{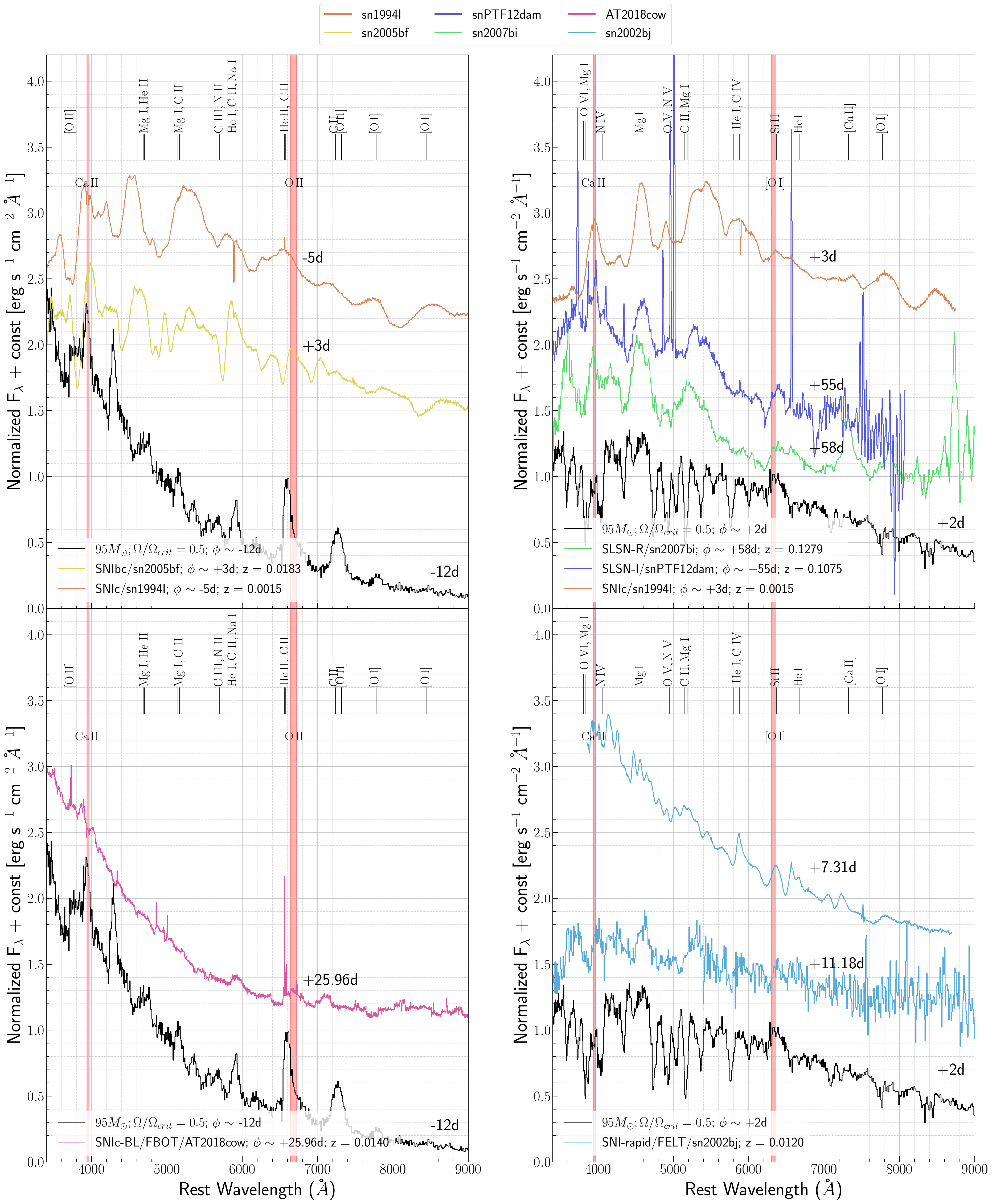}
    \caption{Comparison of the 95 \msun, \rot $= 0.5$ PPI shell collision model spectra with observed SNe best-matched from \texttt{Duperfit} at different epochs where we show the model spectra in black line at $\phi=-12d$ on the first column and at $\phi=+2d$ on the right column. The observed spectra are listed as following. {\it Top Left:} SN Ib/c 2005bf \citep{2006ApJ...641.1039F} and SN Ic 1994I \citep{1995ApJ...450L..11F}. {\it Bottom Left: }FBOT AT2018cow {\it Top Right}: SN Ic 1994I, SLSN-I PTF12dam \citep{2013Natur.502..346N}, and SLSN-R 2007bi \citep{2009Natur.462..624G}. {\it Bottom Right: }FELT SN2002bj. The red bands represent key features: the \caII\ band represents the calcium lines at $\mathrm{\lambda 3939}$ and $\mathrm{\lambda 3969 \text{\AA}}$; the \oII\ band represents the oxygen multiplet at $\mathrm{\lambda 6641}$ and $\mathrm{\lambda 6721 \text{\AA}}$; and the \ooI\ band represents the oxygen forbidden lines at $\mathrm{\lambda 6300}$ and $\mathrm{\lambda 6363 \text{\AA}}$.}\label{fig:spec_day1_day15}
\end{figure*}

To find the best match between our simulated spectrum and SN template spectrum, we apply a range of artificial redshifts ($0.01 \leq z \leq 0.1$ with an increment of 0.01) to our input spectra and perform a redshift grid search in the range of $0.03 \leq z \leq 0.17$ with $\Delta z = 0.01$. Across the fitting, we use the default $R_V$ value, a relative extinction ($-2 \leq A_V \leq 2.0$), the SN template scaling ($0.01 \leq C \leq 3.0$), and the galaxy scaling of $0 \leq D \leq 3$. Figure \ref{fig:spec_day1_day15} shows the best-matched SN spectra to our model spectra at the rest wavelength for different phases ($\phi$), where the phases and redshift values of the observed spectra are displayed on the graphs. 

Although \texttt{Duperfit}'s outputs yield best--fit spectra that are similar to the simulated ones, they are not necessarily best fits, from a quantitative perspective. Our model spectrum on -12d is compared with the spectra of Type Ib/c SN 2005bf (\citealt{2006ApJ...641.1039F}) at $\phi \sim +3$ days, Type Ic SN 1994I (\citealt{1995ApJ...450L..11F}) at $\phi \sim -5$ days, and FBOT AT2018cow at $\phi \sim +25.96$ days. The strong \caII\ emission lines at $\mathrm{\lambda 3939\text{\AA}}$ and $\mathrm{\lambda 3969\text{\AA}}$ and slightly shifted \heII, \cII\ at $\mathrm{\lambda 6563\text{\AA}}$ and $\mathrm{\lambda 6578\text{\AA}}$, respectively, show consistent peaks with SN 2005bf and SN 1994I. In addition, the two \oII\ lines located in the red wing of the broad and asymmetric emission line at $\sim \mathrm{\lambda 6600 \text{\AA}}$  are consistent in both our model and SN 1994I. \citealp{1995ApJ...450L..11F} noted that a possibly present, but weak \heI\ is located at $\mathrm{\lambda 5876\text{\AA}}$ and contaminated by \naI\ $\mathrm{D}$ in SN 1994I, which is also present in our model. Our rotating model further shares a narrows \heI\ emission line at $\mathrm{\lambda 5876\text{\AA}}$ with AT2018cow. At $\mathrm{\lambda 6563\text{\AA}}$, our model exhibits a strong \oII\ emission line while in AT2018cow, there is a nearly featureless He-rich ejecta blend with mininal oxygen signatures, which posts a limitation of our model in explaining the lack of strong oxygen features in AT2018cow. According to \citealt{2018ApJ...865L...3P}, the He emission features in AT2018cow appear redshifted by $\sim 3000$ \kmsone two weeks after maximum, suggesting significant bulk velocity in the \he-rich material.

The later-phase spectrum at +2 days is best matched with SN Ic 1994I at $\phi \sim +3$ days, two luminous SLSNe-I: PTF12dam \citep{2013Natur.502..346N} at $\phi \sim +55$ days and SN 2007bi \citep{2009Natur.462..624G} at $\phi \sim +58$ days, as well as a FELT SN 2002bj at $\phi \sim +7.13 \text{ days and }+11.18$ days. As shown in the top right panel of Figure \ref{fig:spec_day1_day15}, the three observed SN spectra all exhibit a strong \caII\ peak between $\lambda 3939\text{\AA}$ and $\lambda 3969\text{\AA}$, whereas the model spectrum shows only a shallow minimum in that region, although interpretation here is challenging due to the presence of strong line blending. A prominent \mgI\ feature at $\lambda 4560\text{\AA}$ emerges shortly after peak brightness, and forbidden \ooII\ lines blending with \siII\ between $\lambda 6300\text{\AA}$ and $\lambda 6363\text{\AA}$ are evident in all compared SN spectra. Evidence for the presence of silicon in the two \hh-poor shells of the rotating 95 \msun\ model is shown in Figure~\ref{fig:combined_abund}, where the silicon abundance in both pulses is extremely low, with mass fractions ranging from approximately $10^{-9}$ to $10^{-6}$. The spectrum of SN 2002bj, shown in the bottom right panel of Figure \ref{fig:spec_day1_day15}, also features a small, broad feature of \siII, previously aligned with Type Ia SN spectra \citealt{2010Sci...327...58P}. However, SN 2002bj displays prominent \heI\ lines at $\lambda 5876\text{\AA}$ and $\lambda 6678\text{\AA}$ on +7.31 days, unexpected in a Type Ia SN. \cite{2018AAS...23144606R} suggested these \hh\ lines indicate a stripped core-collapse event with a \he-envelope. At $\sim 11$ days, the spectrum of SN 2002bj resembles our model, showing \mgI\ at $\lambda 4571\text{\AA}$, a broad range of absorption features from $\lambda 4600-5200\text{\AA}$, and the presence of \siII\ at $\lambda 6371\text{\AA}$.

The similarities between our model spectra and those of observed Type Ib/c supernovae, suggest that the collision of the two \hh-poor shells produces spectroscopic features broadly consistent with normal \hh-poor explosions, but with some distinctions from the most luminous \hh-poor SLSNe. In specific, while our model captures several key spectroscopic signatures observed in Type Ib/c SNe—such as the emergence of strong $\text{Mg}$ and \oo\ lines and the presence of $\text{Ca}$ features and weak \siII\  features—it does not fully reproduce the persistent, high-luminosity, and flatter light curves characteristic of the most extreme \hh-poor SLSNe like PTF12dam and SN 2007bi. The low silicon content of the ejected shells also affects the strength of certain spectral features, notably weakening \siII\ signatures around $\lambda\,6300$--$6363\,\text{\AA}$, aligning with observed trends in SN 2002bj. Notably, both AT2018cow and SN 2002bj exhibit prominent \heI\ and \heII, aligning with our model’s strong narrow \heI\ and weaker broad \heII\ features. Overall, the spectral evolution of our model reflects the intermediate nature of PPISN shell collisions, producing luminous, rapidly evolving transients with \hh-poor spectra that are fainter and faster-declining than classical SLSNe-I, while sharing notable similarities with fast-evolving events like AT2018cow and SN 2002bj.

\section{DISCUSSION AND CONCLUSIONS} \label{sec:dis}

In this study, we investigated the effects of rotation on PPISN progenitors, focusing on stars with ZAMS masses between 85 and 160 \msun, metallicities of 10\% \zsun and \zsun, and rotation rates ranging from \rot = 0\% to 70\%. Using \texttt{MESA}, we modeled both the evolutionary and dynamical phases of these stars, enabling us to examine the properties of the shell ejections produced during PPISN events. We analyzed the number, strength, and duration of pulses, along with the physical characteristics of each ejected shell, including temperature, density, velocity, and composition. Our results demonstrate that rotation strongly influences the mass-loss history, energetics, and chemical signatures of PPISNe, offering important insights into their role in shaping a range of luminous transients.

Rotation markedly alters PPISN behavior. Higher initial rotation increases wind-driven mass loss during stellar evolution, resulting in lower total ejected mass and reduced kinetic energy in the pulses. This finding aligns with the results of \citet{2012ApJ...748...42C}, who showed that rapid rotation lowers the PI threshold by enhancing internal mixing and mass loss. For example, our 85 \msun\ model at \rot $= 0.7$ ejects only 2.95 \msun\ across three pulses, compared to 7.5 \msun\ at \rot $= 0.2$. This reflects how centrifugal forces and mixing compact the star, suppressing envelope expansion. In contrast, increasing ZAMS mass yields larger CO cores, leading to fewer but more energetic pulses. For instance, the 115 \msun\ model at \rot $= 0.5$ ejects 24 \msun\ across two energetic pulses. These findings are consistent with \citet{2019ApJ...887...72L}, who showed that low metallicity preserves massive CO cores, increasing pulse strength. Our models suggest a key trend: rotation reduces ejected mass, while higher initial mass increases both ejected mass and kinetic energy, with a strong correlation to peak ejecta velocity.

Rotation also significantly impacts chemical evolution, reshaping the composition and dynamics of the PPISN ejecta. Rapidly rotating models produce \hh-poor, \cc- and \oo-rich shells due to enhanced rotational mixing, in agreement with \citet{2012ApJ...760..154C}. For example, in the 95 \msun\ model with \rot $= 0.5$, the second pulse ejects a shell composed of 62\% oxygen and negligible hydrogen, in contrast to the \he-rich first pulse (24\% \he). This reflects chemically homogeneous evolution (CHE), where rotation mixes nuclear-burning products throughout the envelope, resulting in uniform surface enrichment. Enhanced radiatively driven winds—boosted by rotationally increased surface metallicity—strip the envelope prior to the onset of PPISN, as also discussed in \citet{2016MNRAS.457..351Y}.

Metallicity further modulates PPISN outcomes. At solar metallicity, stronger winds reduce CO core masses, causing the 140 \msun\ model to transition from PISN (at 10\% \zsun) to PPISN. This trend, studied by \citet{2019ApJ...882...36M} and \citet{2020A&A...640L..18M}, illustrates how rotation and metallicity jointly shift the PPISN cutoff and expand the range of stars capable of undergoing pulsational pair-instability.

To explore this further, we compared the pulse profiles of rotating PPISN progenitors (85, 90, and 95 \msun) to non-rotating models from \citet{2017ApJ...836..244W} with similar \he-core masses (49-55 \msun). The rotating and non-rotating models produce the same number of pulses and comparable final masses post-PPI. However, non-rotating models like T110B and T123A eject significantly more mass in their first pulse ($\sim 35$ \msun\ and 17.8 \msun, respectively), compared to our rotating counterparts (e.g., $\sim 4$ \msun\ for $110$ \msun\ at \rot $= 0.5$). Nonetheless, our rotating models exhibit higher energy in later pulses with higher velocities (e.g., 4086 \kmsone in the 85 \msun\ model vs. 1300 \kmsone in T110B). Notably, a 90 \msun\ star at \rot $= 0.5$ or $0.7$ forms a 55 \msun\ \he-core that experiences PPI, whereas a non-rotating star requires 123 \msun\ (T123A) to achieve the same. This highlights the role of rotation in lowering the mass threshold for PPISNe. Furthermore, rotating models show no \hh\ and less than 10\% \hh\ in later pulses, producing \hh-poor, \cc- and \oo-rich shells. Their collisions may explain \hh-poor luminous transients such as SLSN-I, Fast Blue Optical Transients (FBOTs), and Fast-Evolving Luminous Transients (FELTs). The shorter interpulse intervals in rotating models (e.g., 26 days for the 95 \msun\ model) enhance the likelihood of shell collisions, contrasting with the century-long delays in non-rotating models like T123A.

To test this scenario, we modeled the light curve and spectra of a two-shell collision for the 95 \msun, \rot $= 0.5$ model using \texttt{STELLA} and \texttt{SuperLite}. The resulting transient reaches a bolometric peak luminosity of $4.8$-$6.3 \times 10^{42}$ ergs$^{-1}$ and remains above $3 \times 10^{40}$ ergs$^{-1}$ for $\sim$50 days—comparable to Type Ib/c SNe like SN 1994I (M$_{\text{B}} \sim -17$), but fainter than SLSNe-I such as PTF12dam (M$_{\text{B}} \sim -22$) and fast-evolving transients like AT2018cow and SN 2002bj. The light curve shows a rapid rise and steep decline, resembling SN 1994I, while the spectral features—broad \cII\ and \oII\ emission before peak and strong \oIV\ and \mgI\ absorption afterward—align with \hh-poor transients such as SN 2005bf and PTF12dam. Our model spectrum also matches the prominent \heI\ and \heII\ lines seen in the FBOT AT2018cow and the FELT SN 2002bj, as well as the weak \siII\ feature in SN 2002bj. However, the light curves of AT2018cow and SN 2002bj evolve more rapidly and luminous than that of our model, suggesting additional mechanisms may drive their fast and bright temporal evolution.  These features are consistent with the predictions from \citet{2016MNRAS.457..351Y} for PPISN-driven Type Ic events. However, the luminosity of the model that was explored in this work is insufficient to explain the most luminous SLSNe, implying that additional power sources such as magnetar spin-down \citep{2010ApJ...719L.204W} or black hole fallback accretion \citep{2013ApJ...772...30D} may be required, particularly for rotating progenitors that retain angular momentum.

The rapid rotation of the iron core in our models raises the possibility of a rapidly rotating remnant black hole, but determining the precise angular velocity of the core is beyond the scope of this study. However, we note that the surface angular velocities at the edge of the iron/silicon core at collapse for models labeled as ``CC'' in Tables \ref{tab:lowZ} and \ref{tab:solarZ} typically range between 200 and 800 \kmsone, which are below the threshold generally associated with collapsar formation \citep{2006ApJ...637..914W}.

Our findings advance the understanding of PPISNe beyond the non-rotating paradigm. Rotation not only reduces ejected mass but also enhances metal enrichment, implying increased injection of synthesized elements into the interstellar medium, potentially influencing early galaxy chemical evolution. The luminous, short-lived shell collisions in rotating models provide a plausible explanation for FBOTs and FELTs, bridging classical core-collapse SNe and SLSNe. This connection has also been explored by \citet{2023MNRAS.526.4130H} in the context of electromagnetic and gravitational-wave transients. Moreover, the compact remnants from our rotating models fall near the lower edge of the black hole mass gap ($\sim$40–50 \msun), suggesting a possible link to the formation of massive black holes and binary BH systems observed by LIGO/Virgo.

However, our models are 1D and assume spherical symmetry, likely underestimating asymmetries such as bipolar outflows. As shown by \citet{2007ApJ...664..416B}, jet-driven explosions and magnetohydrodynamic effects can arise in rapidly rotating progenitors in 2D/3D simulations. We plan to extend this work using 2D and 3D versions of our group's radiation transport code, \texttt{SuperLite}, to model PPISNe in non-spherical geometries. This will enable us to explore how hydrodynamic instabilities and viewing angle affect the resulting light curves and spectra.

In conclusion, our study shows that PPISNe originating from massive rotating stars represent a distinct and significant class of astrophysical transients. Their influence spans from enriching primordial gas in the early Universe to shaping contemporary host galaxy environments and contributing to the formation of massive black holes. These findings validate their potential in explaining \hh-poor, luminous events and highlight rotation as a key parameter in the evolution of massive stars. Future multidimensional modeling of PPISNe will be essential in resolving open questions about their observational diversity and their role in cosmic evolution.

\begin{acknowledgments}
TH would like to take Juhan Frank for useful discussions and Michael Baer for his valuable inputs in \texttt{Duperfit}. EC would like to thank NASA and the Smithsonian Astrophysical Observatory (SAO) for their support via the Chandra X-ray Observatory (CXO) theory grant TM4-25003X.
\end{acknowledgments}

\software{\texttt{MESA} r24.03.1 \citep{2011ApJS..192....3P, 2013ApJS..208....4P, 2015ApJS..220...15P, 2018ApJS..234...34P, 2019ApJS..243...10P}, \texttt{STELLA} \citep{1998ApJ...496..454B, 2000ApJ...532.1132B, 2006A&A...453..229B}, \texttt{SuperLite} \citep{2023ApJ...953..132W, gururaj_wagle_2023_8111119}, \texttt{Duperfit} \citep{2005ApJ...634.1190H, duperfit}}

\bibliography{references}{}
\bibliographystyle{aasjournalv7}



\end{document}